\documentclass[11pt]{article}
\usepackage[T1]{fontenc}
\usepackage{lmodern}
\usepackage[margin=1in]{geometry}
\usepackage{amsmath,amssymb,amsthm,mathtools}
\usepackage[normalem]{ulem}
\usepackage{microtype,booktabs,array,enumitem,xcolor}
\usepackage{tikz}
\usetikzlibrary{arrows.meta,decorations.markings,positioning,calc,fit}
\usepackage[colorlinks=true,linkcolor=blue!55!black,citecolor=blue!55!black,urlcolor=blue!55!black]{hyperref}
\usepackage[nameinlink,noabbrev]{cleveref}
\newtheorem{theorem}{Theorem}[section]
\newtheorem{lemma}[theorem]{Lemma}

\theoremstyle{definition}

\newtheorem{remark}[theorem]{Remark}

\newcommand{\OPT}{\operatorname{OPT}}
\newcommand{\mst}{\operatorname{mst}}
\newcommand{\Oh}{\widetilde O}
\newcommand{\E}{\mathbb E}

\newcounter{algorithm}
\crefname{algorithm}{Algorithm}{Algorithms}
\Crefname{algorithm}{Algorithm}{Algorithms}
\newlist{algorithmsteps}{enumerate}{1}
\setlist[algorithmsteps]{label=\arabic*:,leftmargin=2em,labelsep=0.5em,
  itemsep=1pt,parsep=0pt,topsep=3pt,align=right}
\newenvironment{pseudocode}[1]{%
  \par\addvspace{10pt}\noindent\begin{minipage}{\linewidth}\small
  \refstepcounter{algorithm}\hrule\vspace{4pt}
  \textbf{Algorithm \thealgorithm}\ #1\par\vspace{4pt}\hrule\vspace{2pt}
}{\vspace{3pt}\hrule\end{minipage}\par\addvspace{10pt}}
\setlist{topsep=4pt,itemsep=3pt,parsep=0pt}
\title{Dynamic Steiner Tree under Edge Updates}
\author{
  Jeff Giliberti\thanks{University of Maryland, College Park.
    Emails: \texttt{jeffgili, hajiagha, mahdavi, xcyang@umd.edu}.}
  \qquad
  MohammadTaghi Hajiaghayi\footnotemark[1]
  \\[0.5em]
  Mohammad Mahdavi\footnotemark[1]
  \qquad
  Xinchen Yang\footnotemark[1]
}
\date{}

\begin{document}
\maketitle
\vspace{-2.5em}
\begin{abstract}
The Steiner tree problem is the canonical generalization of minimum spanning tree (MST). When connectivity is required only among a fixed subset of vertices (terminals), the goal is to maintain a minimum-cost tree connecting these terminals as the graph undergoes edge updates. 
While dynamic MST has been studied since the 1980s~[Frederickson, STOC'83, SICOMP'85], and the first sublinear-in-$n$ dynamic MST algorithm dates back almost thirty-five years to Eppstein, Galil, Italiano, and Nissenzweig~[FOCS'92, JACM'97], achieving sublinear-in-$n$ update time for dynamic Steiner tree has long remained open, as a single edge update can invalidate many routes in the underlying terminal metric closure.

We present the first fully dynamic algorithm for Steiner tree that achieves a constant-factor approximation with sublinear-in-$n$ update time. For any fixed $0<\zeta<1/2$, our deterministic algorithm maintains an $O_\zeta(1)$-approximate Steiner tree for undirected graphs with polynomially bounded edge weights in $\widetilde O(n^{1/2+\zeta})$ amortized update time, answering cost queries in $O(1)$ time, and reporting an explicit tree in $\widetilde O(n^{1+\zeta})$ time. Our approach combines periodic terminal-metric rebuilds on a dynamic spanner with a top-tree dynamic program that localizes edge deletions and reconnects surviving components through a dynamic distance sparsifier. Along the way, we give a simpler randomized insertion-only algorithm with a better, explicit approximation ratio.
\end{abstract}

\setcounter{tocdepth}{1}
\tableofcontents
\clearpage

\section{Introduction}
The \textit{Steiner tree problem} asks for a minimum-weight connected subgraph connecting a specified set of terminals in an undirected weighted graph. This NP-hard problem is a fundamental abstraction of network design where a set of target endpoints must be connected with minimum cost using intermediate nodes as optional relays. Consequently, it has motivated decades of work on polynomial-time approximation algorithms in the static setting. A classical algorithm by Kou, Markowsky, and Berman achieved a $2$-approximation by constructing a minimum spanning tree on the metric closure of the terminals (\textit{metric MST})~\cite{KMB81}, and Mehlhorn showed how to compute such a tree in $O(m+n \log n)$ time~\cite{Meh88}. Zelikovsky later improved the approximation factor to $11/6$~\cite{Zel93}, and subsequent work reduced it further to $1.746$~\cite{BR94}, $5/3+\varepsilon$~\cite{PS00}, $1.644$~\cite{KZ97}, and $1.598$~\cite{HP99}. This line of work continued with the $1.55$-approximation of Robins and Zelikovsky~\cite{RZ05}, culminating in the best-known $(\ln 4+\varepsilon)$-approximation of Byrka et al.~\cite{BGRS13} for any fixed $\varepsilon>0$. Our focus is on the \textit{dynamic} version of this problem, where the goal is to maintain an approximate Steiner tree under edge updates, while the vertex and terminal sets remain fixed.

Steiner tree naturally generalizes \textit{minimum spanning tree}: when every vertex is a terminal, the two problems coincide. Dynamic MST is one of the most fundamental problems in dynamic graph algorithms, serving as a natural abstraction of evolving networks and the primary motivation behind foundational dynamic tree data structures and techniques. 
Over forty years ago, Frederickson introduced topology trees to achieve an $O(\sqrt m)$ worst-case update bound for maintaining an MST in an $n$-vertex $m$-edge graph under edge updates~\cite{Fre85}. For the incremental setting, Sleator and Tarjan's link-cut trees~\cite{ST83} maintained an MST in $O(\log n)$ time. To overcome graph density, Eppstein, Galil, Italiano, and Nissenzweig developed sparsification trees to obtain the first bound sublinear in $n$~\cite{EGIN92,EGIN97}. Later, Holm, de Lichtenberg, and Thorup combined Euler tour trees with hierarchical edge levels to achieve $O(\log^4 n)$ amortized update time~\cite{HLT01}. Despite their close relationship and more than four decades of progress on dynamic MST, comparable sublinear update time bounds have not previously been established for constant-factor Steiner tree approximation under edge updates in general weighted graphs. This gap motivates the following question:

\noindent\begin{minipage}{0.99\textwidth}
\bigskip
\begin{center}
    \emph{Is there a fully dynamic (or even incremental) algorithm that maintains an $O(1)$-approximate Steiner tree in general weighted graphs with sublinear-in-$n$ update and cost-query time?}
\end{center}
\smallskip
\end{minipage}

The above question has remained open across almost all regimes, including sparse unweighted graphs. In fact, the only known dynamic algorithm under edge updates, which improves upon recomputing from scratch, is an application of the additive spanner of Bergamaschi, Henzinger, Probst Gutenberg, Vassilevska Williams, and Wein~\cite{BHP21}, which achieves $O(n^{1.529} + k^2 n^{1+o(1)})$ update time, where $k$ is the number of terminals. However, their algorithm applies solely to unweighted graphs, and requires both a sublinear number of terminals and a dense  graph to beat static recomputation. A separate line of work, dating back to Imase and Waxman~\cite{IW91} and studied extensively in online recourse models~\cite{GGK13,GK14} and dynamic settings~\cite{LOP15}, considers terminal arrivals and departures in a \textit{fixed} graph. Leaving the underlying graph metric invariant, their results cannot extend to edge updates. Finally, while dynamic distance oracles and vertex sparsifiers maintain pairwise distances efficiently, they would require at least $\Omega(k)$ work to recalculate  terminal distances. The fundamental obstacle is that a single edge update can simultaneously disrupt many shortest paths shared across multiple terminal pairs, and identifying which routes are invalidated and finding replacement paths in $o(n)$ time has remained unresolved.

\paragraph{Our Contribution.} We give a strong  affirmative answer to the question above. 

\begin{theorem}\label{thm:main-fully-dynamic}
    Let $G$ be a connected $n$-vertex fully dynamic undirected graph with $m_0$ edges initially, a terminal set $R$, and polynomially bounded edge weights over all updates. 
    For every fixed $0<\zeta<1/2$, there is a deterministic fully dynamic algorithm that maintains an $O_\zeta(1)$-approximate Steiner tree for $R$ in $G$ with $\Oh(n^{1/2+\zeta})$ amortized update time and $O((m_0+n)n^\zeta)$ initialization time. 
    After every update, cost queries take $O(1)$ time, and reporting the Steiner tree takes $\Oh(n^{1+\zeta})$ time.
\end{theorem}

Ours is the first fully dynamic algorithm that maintains a constant-factor approximate Steiner tree in sublinear-in-$n$ update time across all graph densities and with no dependence on the terminal-set size. Notably, we also achieve robustness against an adaptive adversary and near-linear $\Oh(n^{1+\zeta})$ reporting time. See also Theorem~\ref{thm:fd-full} for a detailed version of this result.

In terms of techniques, we compute a terminal-metric MST on a sparse dynamic spanner, expanding this MST into a single tree so that any edge deletion breaks at most one tree edge. As an update arrives, a dynamic program over a fixed top-tree hierarchy~\cite{AHLT05} decides in $O(\log n)$ time which parts of this tree to keep, leaving a small number of components to reconnect. We estimate the reconnection cost by running Mehlhorn's algorithm~\cite{Meh88} on a dynamic vertex sparsifier. When an explicit tree is requested, we instead run Mehlhorn's algorithm on the current spanner to reconnect these components. Our final bounds are obtained using the deterministic dynamic spanner of Chuzhoy and Parter~\cite{CP25} and the deterministic dynamic vertex sparsifier of Haeupler, Long, and Saranurak~\cite{HLS24}; any improvement to their results would carry over directly to ours.

When updates consist only of edge insertions, we obtain a simpler randomized algorithm using elementary combinatorial tools that achieves a better approximation constant.

\begin{theorem}\label{thm:main-incremental}
    Let $G$ be a connected $n$-vertex incremental undirected graph with $m_0$ edges initially, a terminal set $R$, and polynomially bounded edge weights over all updates. 
    For every fixed $0<\zeta<1/3$, there is a randomized insertion-only algorithm that maintains an $O(1/\zeta^2)$-approximate Steiner tree for $R$ in $G$ with $\Oh(n^{2/3+\zeta})$ expected amortized update time and $\Oh(m_0+n^{1+\zeta})$ expected initialization time. 
    After every insertion, the approximation guarantee holds deterministically, cost queries take $O(1)$ time, and reporting the Steiner tree takes $\Oh(n^{4/3+\zeta})$ time.
\end{theorem}

In particular, we can get a $(162+\varepsilon)$-approximation in $\Oh(n^{14/15})$ update time; see Theorem~\ref{inc:main} for the full parameterized statement. Our incremental algorithm combines a dynamic spanner~\cite{BKS12} and a static distance oracle~\cite{TZ05} to maintain shortest paths among the endpoints of newly inserted edges. Extending these paths to each endpoint's nearest terminal yields terminal-to-terminal shortcuts whose weights only decrease over time, allowing us to maintain the resulting terminal MST dynamically via link--cut trees~\cite{ST83}.

\subsection{Technical Overview}
\label{sec:overview}
In this work, we develop a dynamic framework that maintains an approximate terminal-metric spanning structure under edge updates, simultaneously achieving a constant approximation factor and sublinear-in-$n$ update time. From this structure we can read off the Steiner tree cost and report its edges efficiently. Our framework is presented in three steps. We begin by examining basic baselines obtained by combining known dynamic data structures and basic Steiner tree algorithms (Section~\ref{sec:overview-reporting-baselines}), showing why off-the-shelf combinations either incur prohibitive query time or sacrifice constant approximation. We then introduce our periodic rebuilding framework in the simpler insertion-only setting (Section~\ref{sec:overview-incremental}), and finally build on these ideas to obtain our deterministic fully dynamic algorithm (Section~\ref{sec:overview-fully-dynamic}).

\subsubsection{Baselines for Steiner cost estimation}
\label{sec:overview-reporting-baselines}
Before describing our framework, we examine how existing dynamic primitives can be used to estimate the Steiner optimum. Table~\ref{tab:related-edge-updates} summarizes the following direct reductions and compares them with our fully dynamic result.

A classical static $2$-approximation offers a natural starting point. For an undirected weighted graph $G$ and terminal set $R$ of size $k=|R|$, consider the complete graph on the terminals denoted by $K_G(R)$, with each edge weighted by its shortest-path distance $d_G(u,v)$ in the underlying graph. An MST of this \emph{terminal metric closure} $K_G(R)$ expands into a Steiner tree of at most twice the optimum weight $\OPT_G(R)$~\cite{KMB81}, cf.\ Lemma~\ref{lem:metric-mst}. To avoid constructing the $\Theta(k^2)$-edge clique $K_G(R)$ explicitly, Mehlhorn~\cite{Meh88} runs a single multi-source Dijkstra computation from $R$ to partition $V$ into Voronoi regions around each vertex's nearest terminal $a(x)\in R$ at distance $\delta(x)=d_G(a(x),x)$, contracts each region into its terminal root so that every cross-region edge $xy$ induces a candidate edge between $a(x)$ and $a(y)$ of weight $\delta(x)+w(xy)+\delta(y)$, and computes an MST of this sparse candidate graph in $O(m+n\log n)$ time, cf.\ Theorem~\ref{thm:mehlhorn}. Maintaining this construction under edge updates to $G_t$, however, presents significant algorithmic challenges. First, a single insertion or deletion can alter shortest-path distances across $\Theta(k^2)$ terminal pairs simultaneously. Second, several barriers emerge when combining this static metric-MST paradigm with known dynamic data structures such as distance oracles, spanners, or tree embeddings.

\paragraph{Distance oracles and vertex sparsifiers.}
A first approach is to use a dynamic distance oracle to estimate pairwise terminal shortest-path distances. For an oracle of stretch $\rho$, querying all $\binom{k}{2}$ terminal pairs reconstructs a $\rho$-approximate terminal metric on $R$. An MST of this graph estimates $\OPT_{G_t}(R)$ within a $2\rho$ factor using $\Oh(k^2)$ distance queries. Plugging in the deterministic dynamic distance oracle of Haeupler, Long, and Saranurak (HLS)~\cite{HLS24}, which for any fixed $0<\eta<1$ achieves stretch $\rho_\eta=2^{\operatorname{poly}(1/\eta)}$, we get $O(n^\eta)$ worst-case update time and an $O(\rho_\eta)$-approximate cost query in $\Oh(k^2)$ time, since each distance query takes $O_\eta(\log\log n)$ time.

To avoid querying all pairs, we can maintain a dynamic vertex sparsifier $H_t$ on the terminal set $R$ that preserves pairwise distances within stretch $\rho$, satisfying $d_{G_t}(u,v)\le d_{H_t}(u,v)\le \rho\,d_{G_t}(u,v)$ for all $u,v\in R$. Running Mehlhorn's linear-time terminal-MST construction directly on $(H_t,R)$ computes an MST of $K_{H_t}(R)$ in $\Oh(|V(H_t)|+|E(H_t)|)$ time, again yielding a $2\rho$-approximate solution. Using the HLS vertex sparsifier~\cite{HLS24}, we can maintain $H_t$ of size $O(kn^\eta)$ in $O(n^\eta)$ amortized update time, reducing the cost-query time to $\Oh(kn^\eta)$.

Both reductions support inexpensive $O(n^\eta)$-time updates to their distance structure, but computing a fresh Steiner cost still requires an MST computation, incurring $\Oh(k^2)$ or $\Oh(kn^\eta)$ time. Moreover, neither structure supports outputting an explicit Steiner tree in $G_t$ directly.

\paragraph{Spanners and Mehlhorn's construction.}
A second approach is to use a dynamic $\sigma$-spanner $S_t\subseteq G_t$, which is a sparse subgraph satisfying $d_{G_t}(u,v)\le d_{S_t}(u,v)\le \sigma\,d_{G_t}(u,v)$ for all vertex pairs. Because $S_t$ is an actual subgraph of $G_t$, running Mehlhorn's construction on $(S_t,R)$ yields an explicit $2\sigma$-approximate Steiner tree in $\Oh(|E(S_t)|+n)$ time. 

Using the deterministic dynamic spanner of Chuzhoy and Parter~\cite{CP25}, which for any fixed $0<\nu<1$ achieves worst-case size $O(n^{1+\nu})$ and stretch $\sigma_\nu=2^{\operatorname{poly}(1/\nu)}$ in $O(n^\nu)$ amortized update time, gives $O(n^\nu)$ update time and $\Oh(n^{1+\nu})$ query time. The main drawback of this approach is that every query requires re-running Mehlhorn's algorithm across all $O(n^{1+\nu})$ spanner edges, even when only an approximate cost is requested. Nonetheless, for reporting an explicit Steiner tree this bound is already near-linear, and our $\Oh(n^{1+\zeta})$ tree-reporting guarantee is achieved by implementing this approach (see Section~\ref{sec:fd-reporting} for details).

\paragraph{Tree embeddings.}
A third approach is to use a dynamic probabilistic dominating tree embedding, which is a rooted tree $T$ with $V(G_t)\subseteq V(T)$ satisfying $d_{G_t}(u,v)\le d_T(u,v)$ for all $u,v$ and expected stretch $\E[d_T(u,v)]\le \alpha\,d_{G_t}(u,v)$ as $G_t$ undergoes edge updates. On a single tree $T$, the minimum-weight subgraph connecting $R$ is simply the unique minimal subtree spanning $R$, of weight $C$. Doubling this subtree and shortcutting between consecutive terminals yields a tour in $K_{G_t}(R)$ of weight at most $2C$, so noncontraction implies $\OPT_{G_t}(R)\le 2C$, while routing an optimal Steiner tree of $G_t$ through $T$ gives $\E[C]\le\alpha\,\OPT_{G_t}(R)$ by linearity of expectation. Augmenting this embedding with top trees~\cite{AHLT05} allows us to maintain the weight $C$ of the minimal subtree spanning $R$, with $O(\log n)$ overhead per change to the embedding, making the $2\alpha$-approximate cost estimate $2C$ available in $O(1)$ query time after every update. 

Plugging in the dynamic probabilistic tree embeddings of Forster, Goranci, and Henzinger (FGH)~\cite{FGH21} yields an $O(\log^5 n)$-approximate cost estimate in expectation with $n^{1/2+o(1)}$ expected amortized update time and $O(1)$ query time. While this combination achieves sublinear update time and $O(1)$ query time for arbitrary $k$, embedding a graph metric into a tree inevitably incurs $\Omega(\log n)$ distortion, so the approximation factor is superconstant and the structure does not provide an explicit Steiner tree.

\begin{table}[htbp]
\centering
\small
\renewcommand{\arraystretch}{1.25}
\setlength{\tabcolsep}{10pt}
\begin{tabular}{@{}llll@{}}
\toprule
Method & Approximation & Update time & Cost query \\
\midrule
HLS oracle + MST
  & $O_\eta(1)$
  & $O(n^\eta)$ worst case
  & $\Oh(k^2)$ \\
HLS sparsifier + Mehlhorn
  & $O_\eta(1)$
  & $O(n^\eta)$ amortized
  & $\Oh(kn^\eta)$ \\
CP spanner + Mehlhorn
  & $O_\nu(1)$
  & $O(n^\nu)$ amortized
  & $\Oh(n^{1+\nu})$ \\
FGH embedding + top trees
  & $O(\log^5 n)$ exp.
  & $n^{1/2+o(1)}$ exp. amortized
  & $O(1)$ \\
\midrule
This work
  & $O_\zeta(1)$
  & $\Oh(n^{1/2+\zeta})$ amortized
  & $O(1)$ \\
\bottomrule
\end{tabular}
\caption{Fully dynamic Steiner tree on $k$ fixed terminals. All rows except FGH are deterministic.}
\label{tab:related-edge-updates}
\end{table}

\subsubsection{An incremental framework}\label{sec:overview-incremental}
The main challenge is to maintain a constant-factor approximation with sublinear update time, independent of the number of terminals. To address it, we develop a framework that combines periodic global rebuilding with efficient local maintenance between rebuilds.

\paragraph{Periodic rebuilding.}\label{inc:periodic-rebuilding}
Our algorithm divides the update sequence into phases of $B$ updates. At the beginning of each phase, we compute and freeze a global approximate solution. After each graph update, we defer global restructuring by combining these frozen structures with Steiner augmentation structures that adapt the maintained solution and preserve a bounded approximation across the phase. If a full global reconstruction incurs a cost of $\mathcal{R}$ and the work for one update within a phase is at most $\mathcal{U}(B)$, the amortized cost per update is bounded by $\mathcal{R}/B + \mathcal{U}(B)$. Choosing the phase length $B$ appropriately balances the heavy cost of periodic rebuilding against the per-update maintenance between rebuilds.

\paragraph{Route maintenance.}\label{inc:phase-init}
At the start of the phase, we take a frozen snapshot $S$ of a dynamic spanner $S_t$~\cite{BKS12} and run Mehlhorn's construction on $(S,R)$ to compute a terminal-metric MST $T_0$, which is a constant approximation at the start of the phase and remains fixed throughout it, along with a static path-reporting distance oracle~\cite{TZ05} on $S$ for querying frozen distances. We also record each vertex's closest terminal $a(x)\in R$, called its owner, and its frozen owner distance $\delta(x)=d_S(a(x),x)$ for later usage. 

As edges are inserted, routes computed at the start of the phase remain valid, but new insertions may create cheaper connections. In the insertion-only setting, any newly useful route must encounter a newly inserted edge and hence one of its endpoints. To capture this, let $F$ denote the set of at most $B$ edges inserted during the phase, and maintain an auxiliary multigraph $L_F$ on the $O(B)$ endpoints of $F$. The edge set of $L_F$ consists of two types of connections between these endpoints: virtual connections supplied by the frozen distance oracle (representing reusable old routes) and newly inserted edges with their actual weights (representing new routing opportunities). We maintain exact all-pairs shortest paths (APSP) on $L_F$ to capture updated routing distances among these endpoints, combining old paths and new insertions. This routing information allows us to adapt the maintained terminal MST by incorporating new connections alongside the fixed baseline $T_0$. 

After each insertion, for each of the $O(B^2)$ endpoint pairs $x,y$ in $L_F$ with distinct owners, we add or update a candidate edge between their owners $a(x)$ and $a(y)$, weighted by $\delta(x)+d_{L_F}(x,y)+\delta(y)$. These candidate weights can only decrease, allowing a link--cut tree to maintain a terminal MST of the augmented graph $T_0\cup C_F$ consisting of $T_0$ and the candidate edges $C_F$ in $\Oh(B^2)$ time per insertion. By the MST augmentation property (Lemma~\ref{lem:mst-augmentation}), augmenting $T_0$ rather than the full frozen terminal metric $K_S(R)$ with $C_F$ preserves the minimum spanning tree weight ($\mst(K_S(R)\cup C_F)=\mst(T_0\cup C_F)$), while nearest-terminal ownership bounds the detour through endpoint owners (Lemma~\ref{inc:sandwich}), ensuring a constant-factor approximation throughout the phase (Theorem~\ref{inc:phase-approximation}). When an explicit Steiner tree is requested, we expand each selected terminal MST edge into its corresponding route in the current graph, using the stored frozen routes, inserted edges, and paths supplied by the distance oracle, and then take their union, remove cycles, and prune nonterminal leaves to obtain the solution. Balancing the $\Oh(n^{1+\zeta}/B)$ amortized phase-rebuild cost against the $\Oh(B^2)$ per-insertion work yields the $\Oh(n^{2/3+\zeta})$ expected amortized update time of Theorem~\ref{thm:main-incremental}.

\subsubsection{Fully dynamic framework}\label{sec:overview-fully-dynamic}
Our fully dynamic algorithm in Section~\ref{sec:fd-design} uses the same periodic rebuilding framework, combining frozen structures with Steiner augmentation structures. Supporting deletions, however, introduces two obstacles to our earlier approach: deleting a single graph edge invalidates all metric edges of $T_0$ whose underlying shortest paths use that edge, and it may destroy frozen routes in $S$ used by the auxiliary multigraph $L_F$. We overcome both obstacles in sublinear deterministic time via four novel ingredients: (1)~expanding $T_0$ into a single tree $P_0$ via coherent tree expansion so that any edge deletion breaks at most one tree edge, (2)~recording both edge insertions and deletions on $P_0$ as zero-weight edges (spokes) to a virtual vertex (hub) in an auxiliary graph $J_t$, (3)~identifying a surviving forest of $O(B)$ components of $P_0$ by maintaining an exact minimum Steiner tree on $J_t$ using dynamic programming on a top tree, and (4)~reconnecting these $O(B)$ surviving components in $o(n)$ time by running Mehlhorn's construction on a dynamic distance sparsifier.

\paragraph{Coherent tree expansion.}
At the beginning of each $B$-update phase on the phase-start graph $G^\circ$, we take a frozen copy $S\subseteq G^\circ$ of the deterministic Chuzhoy--Parter dynamic $\sigma_\nu$-spanner $S_t$~\cite{CP25} and run Mehlhorn's construction on $(S,R)$ to obtain each vertex's nearest terminal owner $a(x)\in R$, its owner distance $\delta(x)=d_S(a(x),x)$, the rooted owner forest, and a terminal-metric MST $T_0$ of $K_S(R)$. Each edge of $T_0$ corresponds to a shortest path in $S$ between two terminals, and several such paths may share edges inside the owner forest. We then apply \emph{coherent tree expansion} (Lemma~\ref{lem:coherent-expansion}) to expand $T_0$ along the rooted owner forest and deduplicate its shared  edges into a single tree $P_0\subseteq S$ spanning $R$ of weight $w(P_0)\le w(T_0)\le 2\sigma_\nu\OPT_{G^\circ}(R)$. Working with $P_0$ rather than $T_0$ localizes structural damage: deleting an edge of $G_t$ removes at most one edge of $P_0$, even if several frozen routes in $T_0$ used it.

\paragraph{Hub-augmented auxiliary graph.}
As updates arrive, deletions may destroy routes in $P_0$, while insertions may create cheaper paths. To adapt $P_0$ to these changes, we construct an auxiliary graph $J_t$ that translates updates into virtual connections. We introduce an artificial vertex $h$, called \emph{hub}, and virtual zero-cost edges, called \textit{spokes}, incident to $h$. At the start of the phase, all edges of $P_0$ are enabled and all spokes are disabled. As updates occur, we update $J_t$ as follows:
\begin{itemize}[nosep]
  \item \emph{Deletion handling:} Deleting $uv\in E(P_0)$ disables $uv$ in $P_0$ and enables the zero-weight spokes $uh$ and $vh$ at its endpoints, providing a zero-cost bridge across the broken route in $P_0$.
  \item \emph{Insertion handling:} Inserting an edge $uv$ enables the zero-weight spokes $a(u)h$ and $a(v)h$ at the endpoints' terminal owners $a(u),a(v)\in R$ on $P_0$, providing a zero-cost connection through $h$ that allows the algorithm to avoid expensive old routes in $P_0$.
\end{itemize}
Letting $D_t\subseteq E(P_0)$ denote the edges of $P_0$ deleted during the phase and $A_t\subseteq R$ denote the \emph{active terminals} (frozen owners of endpoints of newly inserted edges), the set of \emph{active vertices} with enabled spokes is $Z_t=A_t\cup\operatorname{end}(D_t)\subseteq V(P_0)$ of size $O(B)$, and the auxiliary graph is $J_t=(P_0-D_t)\cup\{zh:z\in Z_t\}$.

We maintain an exact minimum Steiner tree $T_t^J$ connecting $R$ in $J_t$ and remove $h$ and its selected spokes. This isolates which parts of $P_0$ are still worth keeping, leaving a forest $F_t=T_t^J-h\subseteq P_0-D_t$ of retained $P_0$ components. Since a phase contains $B$ updates and each update enables at most two spokes, $F_t$ has at most $O(B)$ components, reducing the repair task from connecting all $k$ terminals to connecting at most $O(B)$ components via current distances in $G_t$ in $\Oh(Bn^\eta)$ time. Our analysis bounds the forest's cost $w(F_t)$ and the reconnection cost separately, showing that each is at most a constant multiple of $\OPT_{G_t}(R)$.

To see why $J_t$ always contains a Steiner tree for $R$ of weight at most $2\sigma_\nu\OPT_{G_t}(R)$ (Lemmas~\ref{lem:fd-component} and~\ref{lem:fd-forest-cost}), consider removing all newly inserted edges from a minimum Steiner tree $T^*$ in $G_t$, leaving components $K\subseteq G^\circ$ that consist only of surviving phase-start edges. Rather than doubling all edges of a terminal-containing component $K$ in a closed tour, we traverse $K$ from an endpoint $x\in V(K)$ of a removed newly inserted edge to a terminal $r\in V(K)\cap R$, traversing each edge on the unique $x$--$r$ path in $K$ once and every other edge of $K$ twice, for a walk of length $2w(K)-d_K(x,r)$. Because $a(x)\in A_t$ is the nearest terminal to $x$ in $S$, we have $\delta(x)\le d_S(r,x)\le\sigma_\nu d_K(x,r)$, so the subtraction along the $x$--$r$ walk pays for the shortest path from $a(x)$ to $x$ in $S$. By the MST augmentation property (Lemma~\ref{lem:mst-augmentation}), the terminals of each component $K$ together with $a(x)$ can therefore be connected using routes in the frozen terminal metric $K_S(R)$ of weight at most $2\sigma_\nu w(K)$. Linking these components through their enabled owner spokes $a(x)h$---and bypassing any deleted edge $uv\in D_t$ via its endpoint spokes $u$--$h$--$v$---yields a Steiner tree for $R$ in $J_t$ of weight at most $\sum_K 2\sigma_\nu w(K)\le 2\sigma_\nu\OPT_{G_t}(R)$.

\paragraph{Minimum Steiner tree maintenance on the hub-augmented tree.}
Exact optimization on $J_t$ is enabled by its special ``tree-plus-star'' structure (Section~\ref{sec:optimizer}): its non-spoke edges come from the fixed tree $P_0$, while the virtual spokes form a star centered at $h$. 
At the start of the phase, we build a balanced top-tree hierarchy of clusters~\cite{AHLT05} on the edges of $P_0$ of height $O(\log n)$ exposing a fixed root terminal $r_0\in R$, and run a static dynamic program (DP) to compute the initial optimal solution on $J_t$ with all $P_0$-edges enabled and all spokes disabled. Because each cluster $C$ interacts with the rest of $J_t$ through at most two boundary vertices $\partial C$ in $P_0$ and $h$, each cluster has at most three \emph{interfaces} $I_C=\partial C\cup\{h\}$ and summarizes all partial solutions within $C$ using a constant-size DP table of $22$ states. The top-tree hierarchy built on $P_0$ stays fixed throughout the phase; only the availability of $P_0$-edges and spokes changes due to graph updates. Toggling a $P_0$-edge or spoke updates its assigned cluster's local table and propagates changes up its ancestor path in the balanced hierarchy, recomputing $O(\log n)$ tables in $O(1)$ time each. Since each graph update toggles $O(1)$ edges and spokes, maintaining the DP tables requires only $O(\log n)$ time per update. Because the forest $F_t=T_t^J-h$ may contain $\Theta(n)$ edges, it is represented implicitly by the DP tables and backpointers (Lemma~\ref{opt:dynamic}). We explain next how to use this implicit representation to extract one representative per component of $F_t$ and reconnect those components in $G_t$.

\paragraph{Reconnecting surviving components.}
To reconnect the $O(B)$ components of $F_t$ into a Steiner tree for $R$, we select one \emph{representative} vertex from each component and connect them in the current graph $G_t$ (Section~\ref{sec:repair}). For a candidate subgraph $Y\subseteq J_t$, by lexicographically minimizing \[
\bigl(
w(Y),\,
\text{number of selected spokes},\,
\text{number of selected edges}
\bigr),
\] the DP rules out unnecessary zero-weight spokes and terminal-free components, guaranteeing that every component of $F_t$ contains at least one terminal in $R$ and is incident to a unique selected spoke to $h$ (Lemma~\ref{opt:representatives}). Following the DP backpointers from the root only into clusters with selected spokes, we find the set $Q_t\subseteq Z_t\cup\{r_0\}$ of these $O(B)$ spoke endpoints in $O(B\log n)$ time. Because each component of $F_t$ already connects its terminals to its representative in $Q_t$, connecting $Q_t$ in $G_t$ suffices to connect $R$. Moreover, since every component of $F_t$ contains a terminal in $R$, the union of $F_t$ with an optimal Steiner tree for $R$ in $G_t$ connects all vertices of $Q_t$, proving that $\OPT_{G_t}(Q_t)\le w(F_t)+\OPT_{G_t}(R)\le(1+2\sigma_\nu)\OPT_{G_t}(R)$ (Lemma~\ref{lem:fd-representative-opt}).

To preserve approximate distances among $Q_t$, we maintain the deterministic HLS distance sparsifier $H_t$~\cite{HLS24} (Theorem~\ref{thm:hls}) on the monotonically growing superset $U_t=Z_t\cup\{r_0\}\supseteq Q_t$ of all $O(B)$ active vertices accumulated during the phase. Because $U_t$ gains at most two newly active vertices per graph update and is reset to $\{r_0\}$ at phase boundaries, updating $U_t$ in $H_t$ takes $O(n^\eta)$ amortized time per update and keeps $|V(H_t)|+|E(H_t)|=O(|U_t|n^\eta)=O(Bn^\eta)$ (Lemma~\ref{lem:repair}). By contrast, maintaining $H_t$ on $Q_t$ directly would be slower, as a single edge update can cause the top-tree DP to switch which components connect to the hub and change $\Omega(B)$ vertices of $Q_t$ at once. After each update, we run Mehlhorn's construction on $(H_t,Q_t)$ in $\Oh(Bn^\eta)$ deterministic time to compute a terminal-metric MST on $Q_t$, whose weight (scaled by $\sigma_\nu$) combined with $w(F_t)$ provides our $O(1)$-time cost estimate (Theorem~\ref{thm:fd-phase-approx}). Combining the phase-reconstruction time $\Oh(n^{1+\nu}/B)+O(n^\eta)$ with the single-update processing time $O(n^\nu)+\Oh(Bn^\eta)$ gives amortized update time
\[
T_{\mathrm{upd}}(B)=\Oh\!\left(\frac{n^{1+\nu}}{B}+n^\nu+Bn^\eta\right),
\]
which we balance by setting $B=\bigl\lceil n^{(1+\nu-\eta)/2}\bigr\rceil$ to obtain $\Oh\bigl(n^{(1+\nu+\eta)/2}\bigr)$ deterministic amortized update time, i.e., $\Oh(n^{1/2+\zeta})$ for $\nu=\eta=\zeta$ (Theorem~\ref{thm:main-fully-dynamic}).

\subsection{Related Work} 
We review dynamic Steiner-tree algorithms across several update models, including graph-edge updates and changes to terminals or demands in a fixed metric. We then discuss dynamic graph primitives relevant to our approach and the Steiner cost-estimation bounds that follow from straightforward combinations of these tools.

\paragraph{Steiner trees under graph-edge updates.}
Several algorithms maintain approximate Steiner trees under changes to the underlying graph, providing the closest comparisons to our setting. Bergamaschi et al.~\cite{BHP21} maintain a $(2+\varepsilon)$-approximate Steiner tree in unweighted graphs under both edge and terminal updates, with high probability against an oblivious adversary. For $k$ terminals, their worst-case edge-update time is $O(n^{1.529}+k^2n^{1+o(1)})$, including the time to report an explicit tree. Raikwar and Karmakar~\cite{RK23} give an incremental algorithm for weighted general graphs achieving a $(2-\varepsilon)$-approximation with $O(n)$ update time. Subsequently, Raikwar, Sadharakiya, and Karmakar~\cite{RSK25} give a $(2-\varepsilon)$-approximate incremental algorithm for weighted general graphs with $O(nD_s)$ update time, where $D_s$ is the maximum number of edges on a weighted shortest path. They also give a $(2+\varepsilon)$-approximate fully dynamic algorithm for weighted planar graphs, processing a batch of $b$ updates in $\Oh(k^2\sqrt n+kD+n)$ time for $1\le b\le\sqrt{n+m}$, where $D$ is the unweighted diameter of the graph.

\paragraph{Terminal updates, online recourse, and geometric variants.}
A separate line of work keeps the underlying graph fixed while allowing terminals or connection requirements to change. In a fixed underlying graph, \L{}\k{a}cki et al.~\cite{LOP15} obtain a $(6+\varepsilon)$-approximation with $\Oh(\sqrt n\log D)$ amortized time per terminal insertion or deletion, where $D$ is the metric aspect ratio. Online models often emphasize competitiveness and \emph{recourse}, the number of changes to the maintained solution. This line includes Imase and Waxman~\cite{IW91}, constant-competitive Steiner trees with constant recourse under arrivals~\cite{GGK13}, and constant amortized recourse under terminal arrivals and departures~\cite{GK14}. For Steiner forest, Long et al.~\cite{LMST26} obtain constant competitiveness with logarithmic amortized recourse per arriving demand pair. In fixed-dimensional Euclidean space, Chan et al.~\cite{CGJWX24} maintain an implicit $(1+\varepsilon)$-approximate Steiner tree with polylogarithmic amortized update time for fixed dimension and $\varepsilon$.

\paragraph{Dynamic minimum spanning trees.}
Dynamic MST is the all-terminal special case of our problem and a component of our incremental algorithm. Beyond the classical progression from Frederickson~\cite{Fre85} through sparsification~\cite{EGIN92,EGIN97} to deterministic $O(\log^4 n)$ amortized updates~\cite{HLT01}, Holm, Rotenberg, and Wulff-Nilsen~\cite{HRW15} improved the deterministic amortized bound to $O(\log^4 n/\log\log n)$. For worst-case updates, Kapron, King, and Mountjoy~\cite{KKM13} obtained polylogarithmic Monte Carlo bounds against an oblivious adversary; against an adaptive adversary, Nanongkai, Saranurak, and Wulff-Nilsen~\cite{NSW17} achieved $n^{o(1)}$ worst-case update time w.h.p., Chuzhoy et al.~\cite{CGLNPS20} made this $n^{o(1)}$ worst-case bound deterministic, and Meierhans, Probst Gutenberg, and Yeh~\cite{MPY26} recently achieved polylogarithmic worst-case update time w.h.p. 

\paragraph{Dynamic spanners.}
Dynamic spanners reduce the cost of our periodic rebuilds by preserving approximate distances in sparse subgraphs~\cite{PS89,ADDJS93,BS07}. Following early dynamic constructions~\cite{ADFIR07,Elk07}, Baswana, Khurana, and Sarkar~\cite{BKS12} obtained $(2s-1)$-spanners of size $\Oh(n^{1+1/s})$ in polylogarithmic expected amortized update time against an oblivious adversary, which we combine with geometric weight bucketing in our insertion-only algorithm. Subsequent work achieved faster amortized maintenance~\cite{FG19}, sublinear worst-case update time~\cite{BK16,BFH21}, and adaptive-adversary or deterministic guarantees~\cite{BvdBGNSSS22,BSS22,vdBCKLPGSS23}, culminating in the deterministic constant-stretch spanners of Chuzhoy and Parter~\cite{CP25} via low-diameter router decompositions, which we use in our fully dynamic algorithm. Preserving distances alone, however, does not maintain a low-cost connection among the terminals; our use of sparsification is therefore coupled with periodic rebuilding and an optimizer for the retained solution.

\paragraph{Distance oracles and terminal sparsifiers.}
Distance oracles and terminal sparsifiers supply the distance and path information used to construct and repair terminal connections in our algorithms. Thorup--Zwick oracles~\cite{TZ05} supply the static distance interface in our incremental algorithm. Fully dynamic constant-stretch oracles include the randomized construction of Forster et al.~\cite{FGNS23} and the deterministic construction of Haeupler, Long, and Saranurak (HLS)~\cite{HLS24}. For fixed $\eta>0$, HLS achieve stretch $A_\eta=2^{\operatorname{poly}(1/\eta)}$, $O(n^\eta)$ worst-case update time, and $O_\eta(\log\log n)$ distance-query time. Their separate terminal-sparsifier maintains a graph $H$ of size $O(kn^\eta)$ preserving terminal distances within $A_\eta$, in $O(n^\eta)$ \emph{amortized} update time. Kyng, Meierhans, and Probst Gutenberg~\cite{KMG24} give a complementary deterministic construction with subpolynomial stretch and worst-case update time, together with low-congestion embeddings. 

\paragraph{Tree embeddings and cost estimates.}\label{inc:tree-embeddings}
Dynamic tree embeddings provide another route to Steiner cost estimation by approximating a graph's metric with a tree metric. While probabilistic tree embeddings were pioneered in the static setting by Bartal~\cite{Bar96, Bar98} and subsequently refined with optimal $O(\log n)$ distortion bounds by Fakcharoenphol, Rao, and Talwar~\cite{FRT04}, maintaining such metric approximations dynamically under edge updates has long proved challenging. Forster, Goranci, and Henzinger (FGH)~\cite{FGH21} advanced this frontier by designing dynamic algorithms for maintaining low-stretch probabilistic tree embeddings under edge updates, applying them to problems such as buy-at-bulk network design. Nevertheless, while black-box tree embeddings offer flexible metric reductions, translating tree-based solutions back into valid graph paths while guaranteeing sublinear update times and tight constant-factor approximations for fully dynamic Steiner tree remains difficult, motivating our direct structural and sparsification-based approach.

\paragraph{Organization.}
Section~\ref{sec:preliminaries} presents the model and assumptions, algorithmic primitives, and terminal-metric MST tools. Section~\ref{sec:insertion-only} introduces the insertion-only framework as a warm-up, with implementation details and analysis deferred to Appendix~\ref{app:insertion-only}. Section~\ref{sec:fd-design} develops the fully dynamic algorithm and establishes its approximation, time, and space guarantees.

\section{Preliminaries}
\label{sec:preliminaries}

We first specify the graph model, weight assumptions, and how the maintained solution is represented. We then state the external spanner and distance primitives used by our algorithms. Section~\ref{sec:terminal-mst} develops the terminal-MST construction and its properties, used by both algorithms.

\subsection{Model and Assumptions}
\label{sec:model}

\paragraph{Graph and update model.}
Let $G_t=(V,R,E_t,w_t)$ be an undirected graph on a fixed vertex set $V$, $|V|=n$, with a fixed terminal set $R\subseteq V$, $|R|=k\ge2$. All edge weights are strictly positive. Write $m_0$ for the initial edge count and $m_t$ for the current edge count; $m$ is an upper bound on the total edge count throughout. We assume that the update sequence has polynomial length in $n$.

The fully dynamic algorithm supports edge insertions, deletions, and weight changes (a weight change is processed internally by deleting the old edge and inserting a new edge with the updated weight, without affecting correctness or bounds), and is deterministic (thus holding against an \emph{adaptive adversary}). The incremental algorithm supports only edge insertions and is randomized, with amortized update time bounded in expectation against an \emph{oblivious adversary}. We assume that the initial graph and the graph resulting from each update are \textit{connected}. Both algorithms always return a valid solution with the stated approximation guarantee. The notation $\Oh$ suppresses polylogarithmic factors in $n$. 

\paragraph{Weights.}
We assume that all input weights throughout the update sequence lie in $[1,W]$, where $W=n^{O(1)}$. For some $0<\varepsilon<1$, the fully dynamic algorithm uses rounded edge weights
\begin{equation}\label{eq:rounding}
\widehat w(e)=\left\lceil\frac{w(e)}{\varepsilon}\right\rceil.
\end{equation}
The effect of this rounding on the approximation ratio is analyzed in Lemma~\ref{lem:rounding}.

\paragraph{Notation.}
For a weighted graph $Q$, let $d_Q$ denote shortest-path distance and let $\OPT_Q(U)$ be the minimum weight of a connected subgraph containing a specified vertex set $U$. With nonnegative weights, cycles can be removed without increasing the cost, so $\OPT_Q(U)$ is also the minimum weight of a Steiner tree connecting $U$. We write $\mst(Q)$ for the weight of a minimum spanning tree of $Q$.

The \emph{terminal metric closure} $K_Q(U)$ is the complete graph on $U$ with edge weight $d_Q(u,v)$. For a tree $T$ on $U$, we write
\[
w_{K_Q(U)}(T)=\sum_{uv\in E(T)}d_Q(u,v).
\]

\paragraph{Maintained Output.}
After each update, our algorithms maintain the approximate Steiner tree implicitly by storing a compact representation of the solution (along with its cost estimation, which can be read in $O(1)$ time) that, with additional reporting time, can be expanded to retrieve an explicit  Steiner tree in $G_t$. Our update-time bounds refer to maintaining the implicit solution. Reporting a solution explicitly incurs additional time that is analyzed separately. Note that the cost of reporting the approximate solution can be $\Omega(n)$, since an optimal Steiner tree may contain $\Theta(n)$ edges.

\subsection{Algorithmic Primitives}
We introduce several dynamic graph primitives used by our algorithms and adapt their interfaces to our setting. Our insertion-only algorithm pairs BKS's randomized fully dynamic weighted spanner~\cite{BKS12} with the static path-reporting distance oracle of Thorup and Zwick~\cite{TZ05}, while our fully dynamic algorithm pairs the deterministic dynamic spanner of Chuzhoy and Parter~\cite{CP25} with the deterministic dynamic distance sparsifier of Haeupler, Long, and Saranurak~\cite{HLS24}.

\paragraph{Fully dynamic weighted spanners.}
\label{sec:weighted-spanner}
A spanner is a subgraph that approximately preserves distances in the original graph. In our insertion-only algorithm, we use a weighted extension of BKS's randomized spanner~\cite{BKS12} to maintain a sparse graph under edge updates.

\begin{theorem}\label{thm:dynamic-spanner}
Fix an integer $s\ge2$ and a constant $0<\delta<1$. For undirected graphs with a polynomially bounded edge-weight ratio over the entire update sequence, a randomized fully dynamic structure maintains an explicit subgraph $S_t\subseteq G_t$ such that
\[
d_{G_t}(u,v)\le d_{S_t}(u,v)
\le\sigma d_{G_t}(u,v),
\qquad \sigma=(2s-1)(1+\delta).
\]
The expected size is $\Oh(sn^{1+1/s})$, and the expected amortized update time is $\Oh(s^2)$ against an oblivious update sequence. The stretch guarantee holds deterministically. Starting from $m_0$ edges, initialization takes $\Oh(n+s^2m_0)$ time. 
\end{theorem}
\begin{proof}[Proof Sketch.]
 For unweighted graphs, \cite{BKS12} maintains a $(2s-1)$-spanner of expected size $O(sn^{1+1/s}\log n)$ in $O(s^2\log^2 n)$ expected amortized time per edge insertion or deletion. Following~\cite[Remark~1.2]{BKS12}, we partition the edges into $O(\log_{1+\delta}(w_{\max}/w_{\min}))=O(\log n)$ weight buckets with ratio $1+\delta$ and maintain an unweighted $(2s-1)$-spanner within each bucket; their union achieves weighted stretch $(2s-1)(1+\delta)$ at the cost of an $O(\log n)$ factor in size and update time, absorbed in $\Oh$. Initializing the $O(\log n)$ empty bucket structures on $n$ vertices and inserting the $m_0$ initial edges takes expected $\Oh(n+s^2m_0)$ time, and weight changes are handled by deleting and reinserting the edge.
\end{proof}

For our fully dynamic algorithm, we use the deterministic weighted spanner of Chuzhoy and Parter~\cite{CP25}.

\begin{theorem}\label{thm:cp-spanner}
For any fixed $0<\nu<1$ and an undirected $n$-vertex graph $G_t$ with polynomially bounded positive integer edge weights over the entire update sequence, a deterministic fully dynamic data structure maintains an explicit subgraph $S_t\subseteq G_t$ satisfying
\[
  d_{G_t}(u,v)\le d_{S_t}(u,v)\le\sigma_\nu d_{G_t}(u,v),
\]
with worst-case size $|E(S_t)|=O(n^{1+\nu})$, amortized update time $O(n^\nu)$, initialization time $O((m_0+n)n^\nu)$ from $m_0$ initial edges, and space $O(m_t n^\nu + n^{1+\nu})$, where $m_t=|E(G_t)|$ and the stretch $\sigma_\nu=2^{\operatorname{poly}(1/\nu)}$ depends only on $\nu$.
\end{theorem}
\begin{proof}[Proof Sketch]
By~\cite[Corollary~9.2]{CP25}, for integer parameters $512\le k\le(\log n)^{1/49}$ and $1/\delta_{\mathrm{CP}}\in[400,k]$, an initially empty $n$-vertex simple graph undergoing at most $n^2$ edge insertions and deletions with integer weights in $\{1,\dots,W'\}$ admits a deterministic dynamic spanner of stretch $k^{19}\cdot 2^{O(1/\delta_{\mathrm{CP}}^6)}$ and worst-case size $O(n^{1+O(1/k)}\log W')$ in $n^{O(\delta_{\mathrm{CP}})}$ worst-case update time. Given a fixed $0<\nu<1$, we choose integers $1/\delta_{\mathrm{CP}}=\Theta(1/\nu)\ge 400$ and $k=\max\{1/\delta_{\mathrm{CP}},\Theta(1/\nu)\}\ge 512$ sufficiently large so that the $n^{O(\delta_{\mathrm{CP}})}$ update time is $O(n^\nu)$ and, since $\log W'=O(\log n)$, the $O(n^{1+O(1/k)}\log W')$ size bound is $O(n^{1+\nu/2}\log n)=O(n^{1+\nu})$, giving $\sigma_\nu=2^{O(1/\nu^6)}$. Re-initializing the structure from scratch every $\max\{\lceil m'/4\rceil,n\}$ updates (where $m'$ is the edge count at the last rebuild, so at most $m'+2\max\{\lceil m'/4\rceil,n\}<n^2$ edge operations occur between rebuilds) extends the guarantee to arbitrary $m_0$ and polynomial-length update sequences with $O(n^\nu)$ amortized update overhead and $O(m_t n^\nu+n^{1+\nu})$ space.
\end{proof}

\paragraph{Static path-reporting distance oracle.}
Our insertion-only algorithm uses the static path-reporting distance oracle of Thorup and Zwick~\cite{TZ05}:

\begin{theorem}[\cite{TZ05}]\label{inc:oracle}
For an integer $r\ge2$, an undirected weighted $n$-vertex graph $G$ with $M$ edges admits expected preprocessing time $\Oh(rMn^{1/r})$ and expected space $\Oh(rn^{1+1/r})$. For any connected pair $x,y$, a query takes $\Oh(r)$ time and returns an estimate satisfying
\[
d_G(x,y)\le\widehat d_G(x,y)\le(2r-1)d_G(x,y).
\]
On request, it reports a simple path of weight at most this estimate in $\Oh(r+\ell)$ time, where $\ell$ is the number of reported edges.
\end{theorem}

\paragraph{HLS distance sparsifiers.}
\label{sec:hls-primitive}
We use the dynamic distance sparsifier of Haeupler, Long, and Saranurak~\cite{HLS24}---a smaller weighted graph that preserves approximate distances among vertices from a designated set. Unlike a spanner, this object need not be a subgraph of the original graph, and it may contain auxiliary vertices and edges. Let us first recall their result. We write $\theta$ for their parameter $\epsilon$.

\begin{theorem}[{\cite[Theorem~13.1(1)]{HLS24}}]
\label{thm:hls-original}
For some absolute constant $c>0$, any parameter $1/\log^c n\le\theta\le1$, and an undirected $n$-vertex graph with polynomially bounded positive integer weights, there is a deterministic data structure maintaining an explicit sparsifier $H_t$ for a designated vertex set $U_t$, satisfying
\[
\frac{d_{G_t}(u,v)}{\alpha_{\mathrm{low}}(\theta)}
\le d_{H_t}(u,v)
\le \alpha_{\mathrm{up}}(\theta)d_{G_t}(u,v)
\qquad(u,v\in U_t),
\]
with size $|U_t|n^{O(\theta^4)}$, amortized update time
$n^{O(\theta)}$ per edge or designated-vertex insertion or deletion, and initialization time $(m_0+n)n^{O(\theta)}$, where the approximation factors $\alpha_{\mathrm{low}}(\theta),\alpha_{\mathrm{up}}(\theta)=2^{\operatorname{poly}(1/\theta)}$ depend only on $\theta$.
\end{theorem}

We make some slight modifications to their result to simplify its later usage. 

\begin{theorem}[cf. \cite{HLS24}]\label{thm:hls}
For any fixed $0<\eta<1$ and an undirected $n$-vertex graph $G_t$ with polynomially bounded positive integer weights, there is a deterministic data structure maintaining an explicit sparsifier $H_t$ with positive edge weights for a designated vertex set $U_t\subseteq V$, satisfying
\[
d_{G_t}(u,v)\le d_{H_t}(u,v)\le\rho_\eta d_{G_t}(u,v)
\qquad(u,v\in U_t),
\]
with size $|V(H_t)|+|E(H_t)|=O(|U_t|n^\eta)$, amortized update time $O(n^\eta)$ per edge or designated-vertex insertion or deletion, initialization time $O((m_0+n)n^\eta)$, and space $O((m_t+n)n^\eta)$, where $m_t=|E(G_t)|$ and the approximation factor $\rho_\eta=2^{\operatorname{poly}(1/\eta)}$ depends only on $\eta$.
\end{theorem}
\begin{proof}[Proof Sketch]
Given a fixed $0<\eta<1$, we choose $\theta=\Theta(\eta)\in(0,1)$ sufficiently small so that the $n^{O(\theta^4)}$ and $n^{O(\theta)}$ factors in Theorem~\ref{thm:hls-original} are bounded by $O(n^\eta)$. Scaling all edge weights of $H_t$ by $\alpha_{\mathrm{low}}(\theta)$ ensures noncontraction ($d_{G_t}(u,v)\le d_{H_t}(u,v)\le\rho_\eta d_{G_t}(u,v)$) with $\rho_\eta=\alpha_{\mathrm{low}}(\theta)\alpha_{\mathrm{up}}(\theta)=2^{\operatorname{poly}(1/\eta)}$. Re-initializing the structure from scratch on $(G_t,U_t)$ every $L=\max\{\lceil m'/4\rceil,n\}$ operations (where $m'$ is the edge count at initialization or the last rebuild, so $m'+n+L=O(m_t+n)$ throughout the window and the $O((m'+L+n)n^\eta)=O(Ln^\eta)$ rebuild cost charges $O(n^\eta)$ per operation) bounds the total work---and hence space---of the active instance by $O((m_t+n)n^\eta)$.
\end{proof}

\subsection{Terminal-Metric MSTs}
\label{sec:terminal-mst}
We introduce several basic concepts and constructions used to approximate the Steiner tree. A classical approach to approximating the Steiner tree~\cite{KMB81} reduces finding a Steiner tree for a terminal set $U$ in $Q$ to computing an MST of the terminal metric closure $K_Q(U)$ and unfolding its edges into shortest paths in $Q$, losing at most a factor of $2$.
 
\begin{lemma}[Metric MST]\label{lem:metric-mst}
For a connected graph $Q$ with positive edge weights and a nonempty terminal set $U\subseteq V(Q)$,
\[
\mst(K_Q(U))\le2\OPT_Q(U),
\]
that is, the weight of an MST of the terminal metric closure is at most twice the weight of an optimal Steiner tree connecting $U$ in $Q$. Moreover, replacing each edge of a spanning tree of $K_Q(U)$ by a shortest path in $Q$ and taking the union of these paths yields a connected subgraph containing $U$, whose weight is no greater than that of the spanning tree. Deleting cycles and pruning leaves outside $U$ yields a $2$-approximate Steiner tree.
\end{lemma}
\begin{proof}
 Doubling the edges of an optimal Steiner tree for $U$ in $Q$ yields an Eulerian tour of weight $2\OPT_Q(U)$ visiting all terminals in $U$. Shortcutting this tour between consecutive first visits to terminals in $U$ gives a spanning path of $K_Q(U)$ of weight at most $2\OPT_Q(U)$ by the triangle inequality, so $\mst(K_Q(U))\le 2\OPT_Q(U)$. Expanding the edges of any spanning tree of $K_Q(U)$ into shortest paths in $Q$ and taking their union connects $U$ while counting shared edges only once, and removing cycles and non-$U$ leaves preserves connectivity on $U$ without increasing weight.
\end{proof}

\paragraph{Nearest-terminal forest.}
To avoid constructing the $\Theta(|U|^2)$-edge metric clique $K_Q(U)$ explicitly, Mehlhorn~\cite{Meh88} partitions $V(Q)$ into Voronoi regions around the terminals. Given a connected graph $Q$ with $N$ vertices, $M$ edges, and positive edge weights, and a nonempty terminal set $U\subseteq V(Q)$, a single multi-source Dijkstra computation initialized with all terminals in $U$ (with consistent tie-breaking) runs in $O((M+N)\log N)$ time and assigns each vertex $x\in V(Q)$ a nearest terminal $a(x)\in U$, called its \emph{owner}, at distance
\[
\delta(x)=d_Q(a(x),x)=\min_{u\in U}d_Q(u,x).
\]
The shortest-path tree edges recorded by Dijkstra form a rooted forest---called the \emph{owner forest}---in which each terminal $u\in U$ is the root of a tree spanning its owned region $\{x\in V(Q):a(x)=u\}$.

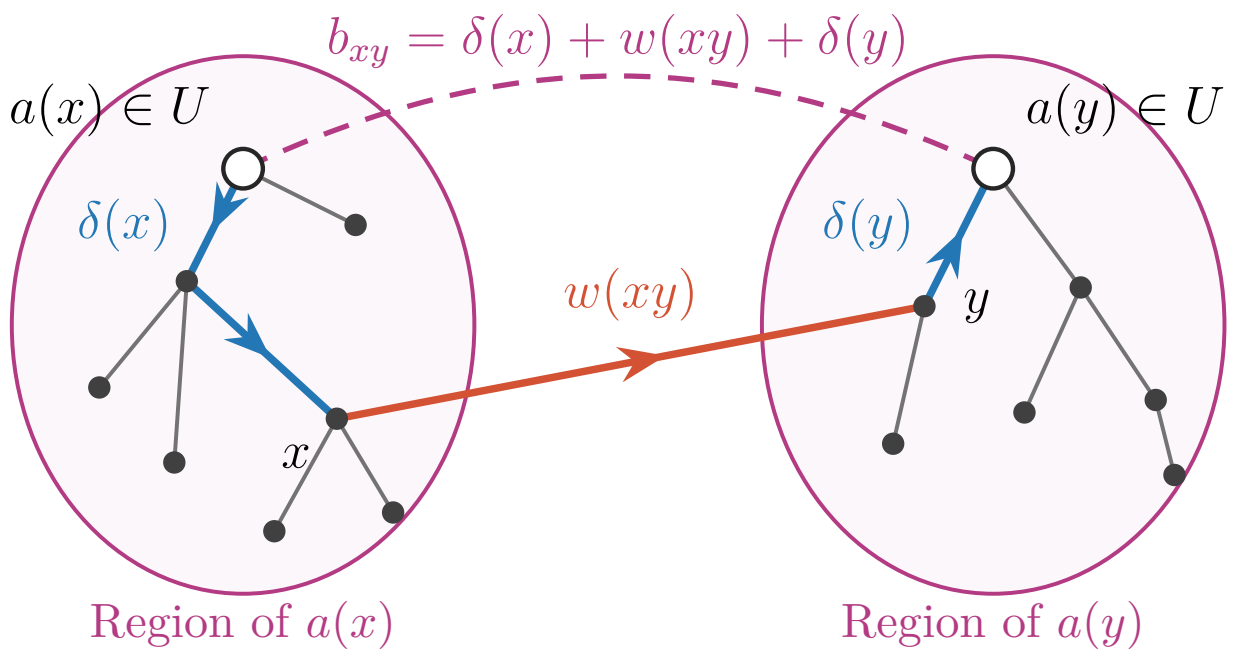
\begin{figure}[ht]
\centering
\resizebox{0.5\linewidth}{!}{%
\begin{tikzpicture}[
    vertex/.style={circle,fill=black!75,inner sep=1.8pt},
    terminal/.style={circle,draw=black!85,fill=white,line width=1pt,minimum size=9pt,inner sep=0pt},
    tree edge/.style={draw=black!55,line width=0.9pt},
    route/.style={line width=1.7pt,postaction={decorate},decoration={markings,mark=at position 0.56 with {\arrow{Stealth}}}},
    region/.style={draw=regioncolor,fill=regioncolor!4,line width=0.9pt}
]
\definecolor{regioncolor}{HTML}{B33B83}
\definecolor{routecolor}{HTML}{2377B5}
\definecolor{crosscolor}{HTML}{D45234}

\draw[region] (2,1.85) ellipse [x radius=1.85,y radius=2.15];
\draw[region] (8,1.85) ellipse [x radius=1.85,y radius=2.15];

\coordinate (ax) at (2,3.1);
\coordinate (l1) at (1.55,2.2);
\coordinate (x) at (2.75,1.1);
\coordinate (l2) at (0.85,1.35);
\coordinate (l3) at (1.45,0.75);
\coordinate (l4) at (2.25,0.2);
\coordinate (l5) at (3.2,0.35);
\coordinate (l6) at (2.9,2.65);

\coordinate (ay) at (8,3.1);
\coordinate (y) at (7.45,2);
\coordinate (r1) at (7.2,0.9);
\coordinate (r2) at (8.7,2.15);
\coordinate (r3) at (8.25,1.15);
\coordinate (r4) at (9.3,1.25);
\coordinate (r5) at (9.45,0.65);

\draw[tree edge] (ax)--(l6);
\draw[tree edge] (l1)--(l2);
\draw[tree edge] (l1)--(l3);
\draw[tree edge] (x)--(l4);
\draw[tree edge] (x)--(l5);
\draw[tree edge] (y)--(r1);
\draw[tree edge] (ay)--(r2);
\draw[tree edge] (r2)--(r3);
\draw[tree edge] (r2)--(r4);
\draw[tree edge] (r4)--(r5);

\draw[route,draw=routecolor] (ax)--(l1);
\draw[route,draw=routecolor] (l1)--(x);
\draw[route,draw=crosscolor] (x)--(y);
\draw[route,draw=routecolor] (y)--(ay);

\draw[draw=regioncolor,line width=1.2pt,dash pattern=on 6pt off 4pt] (ax) to[out=25,in=155] (ay);
\node[text=regioncolor,font=\large] at (5,4.12) {$b_{xy}=\delta(x)+w(xy)+\delta(y)$};

\foreach \p in {l1,x,l2,l3,l4,l5,l6,y,r1,r2,r3,r4,r5}
    \node[vertex] at (\p) {};
\node[terminal] at (ax) {};
\node[terminal] at (ay) {};

\node[above left=5pt,font=\large] at (ax) {$a(x)\in U$};
\node[above right=5pt,font=\large] at (ay) {$a(y)\in U$};
\node[below left=3pt,font=\large] at (x) {$x$};
\node[right=5pt,font=\large] at (y) {$y$};
\node[text=routecolor,font=\large] at (1.05,2.65) {$\delta(x)$};
\node[text=routecolor,font=\large] at (7.0,2.65) {$\delta(y)$};
\node[text=crosscolor,font=\large,above] at (5.1,1.75) {$w(xy)$};
\node[text=regioncolor] at (2,-0.55) {Region of $a(x)$};
\node[text=regioncolor] at (8,-0.55) {Region of $a(y)$};
\end{tikzpicture}%
}
\caption{A cross-region edge $xy$ induces a candidate edge between terminals $a(x)$ and $a(y)$ with weight $b_{xy}=\delta(x)+w(xy)+\delta(y)$. Arrows indicate its route in $Q$ from $a(x)$ to $a(y)$; the dashed curve represents the candidate edge.}
\label{fig:mehlhorn-construction}
\end{figure}

\paragraph{Cross-region edges and metric MST construction.}
Given the partition of $V(Q)$ into terminal owner regions, we can replace the dense metric closure $K_Q(U)$ by a sparse graph on $U$ generated only by edges that cross between different regions.
Call an edge $xy$ a \emph{cross-region edge} if its endpoints belong to different owner regions, namely, $a(x)\ne a(y)$. Conceptually, represent each region by a supernode identified with its terminal root. Each cross-region edge $xy$ gives a candidate edge between terminals $a(x)$ and $a(y)$ with weight
\begin{equation}\label{eq:boundary-candidate}
b_{xy}=\delta(x)+w(xy)+\delta(y).
\end{equation}
This weight is the length of a realizable route in $Q$ that follows the owner tree from $a(x)$ to $x$, crosses $xy$, and follows the other owner tree from $y$ to $a(y)$. Figure~\ref{fig:mehlhorn-construction} illustrates this construction.

Construct the candidate graph $C_0$ by scanning all $M$ edges of $Q$ and discarding those whose endpoints have the same owner. Among all candidate edges crossing the same two regions, retain one candidate edge of minimum weight. The resulting graph $C_0$ is a simple weighted graph on $U$ with at most $M$ edges.

\begin{lemma}[Equality of cut minima]\label{lem:cut-minima}
For any partition of terminal set $U$ into two nonempty groups $A$ and $U\setminus A$, the cheapest edge crossing this cut has the same weight in $C_0$ and in $K_Q(U)$.
\end{lemma}
\begin{proof}
Every candidate edge in $C_0$ represents a walk in $Q$ between its terminal endpoints, so its weight is at least their distance in $Q$; hence the minimum weight crossing $(A,U\setminus A)$ in $C_0$ is at least that in $K_Q(U)$. Conversely, let $u\in A$ and $v\in U\setminus A$ minimize $d_Q(u,v)$, and fix a shortest $u$--$v$ path $P$ in $Q$. Since $a(u)=u\in A$ and $a(v)=v\in U\setminus A$, the path $P$ contains an edge $xy$ with $a(x)\in A$ and $a(y)\in U\setminus A$. Because $\delta(x)\le d_Q(u,x)$ and $\delta(y)\le d_Q(y,v)$, the induced candidate edge between $a(x)$ and $a(y)$ crosses $(A,U\setminus A)$ with weight
\[
b_{xy}=\delta(x)+w(xy)+\delta(y)\le d_Q(u,x)+w(xy)+d_Q(y,v)=d_Q(u,v),
\]
and the lightest parallel candidate edge retained in $C_0$ is no heavier.
\end{proof}

Because $C_0$ and $K_Q(U)$ have identical minimum cut weights across every partition of $U$, the cut property of minimum spanning trees implies that an MST of the sparse candidate graph $C_0$ is also an exact MST of $K_Q(U)$.

\begin{theorem}[Mehlhorn terminal-MST construction~\cite{Meh88}]\label{thm:mehlhorn}
An MST $T_0$ of the candidate graph $C_0$ is also an MST of the terminal metric closure $K_Q(U)$. Every selected candidate edge has weight equal to the distance in $Q$ between its terminal endpoints and therefore represents an exact shortest path between them. The entire construction takes $\Oh(M+N)$ time and $O(M+N)$ working space, without explicitly constructing the metric clique.
\end{theorem}\begin{proof}
Starting from an MST of $K_Q(U)$, repeatedly replacing each metric edge by a lightest edge in $C_0$ crossing the induced cut (which has no greater weight by Lemma~\ref{lem:cut-minima}) produces a spanning tree of $C_0$ of weight at most $\mst(K_Q(U))$, so $\mst(C_0)\le\mst(K_Q(U))$. Conversely, every edge $uv\in E(C_0)$ corresponds to a walk in $Q$ of weight $w_{C_0}(uv)\ge d_Q(u,v)$, so for an MST $T_0$ of $C_0$,
\[
\mst(K_Q(U))\le\sum_{uv\in E(T_0)}d_Q(u,v)\le\sum_{uv\in E(T_0)}w_{C_0}(uv)=\mst(C_0)\le\mst(K_Q(U)).
\]
Equality holds throughout, so $T_0$ is an MST of $K_Q(U)$ and $w_{C_0}(uv)=d_Q(u,v)$ for every $uv\in E(T_0)$. Finally, running multi-source Dijkstra from $U$, scanning the $M$ edges of $Q$ to build and deduplicate $C_0$, and computing an MST of $C_0$ (on $|U|\le N$ vertices and at most $M$ edges) each take $\Oh(M+N)$ time and $O(M+N)$ space.
\end{proof}

\paragraph{Coherent tree expansion.}
Each edge of the terminal-metric MST $T_0$ represents a shortest path in $Q$ between two terminals, and several such paths may share edges inside the owner forest. In the dynamic setting, maintaining these paths separately would allow a single edge deletion to invalidate many routes of $T_0$ simultaneously. Instead, we expand $T_0$ along the rooted owner forest and deduplicate shared predecessor edges to obtain a single tree $P_0\subseteq Q$ spanning $U$, localizing the effect of any edge deletion to a single edge of $P_0$.

\begin{lemma}[Coherent expansion guarantee]\label{lem:coherent-expansion}
The union $P_0$ obtained by expanding the terminal-metric MST $T_0$ using the owner forest and the corresponding cross-region edges is a tree in $Q$ containing the terminal set $U$, and
\[
w(P_0)\le w(T_0)\le2\OPT_Q(U).
\]
Given the owner forest and these cross-region edges, $P_0$ can be constructed in $O(N)$ time.
\end{lemma}
\begin{proof}
Within each owner region, the union of the owner-forest paths from the endpoints of selected cross-region edges to their terminal root forms a subtree of that region's owner tree. Contracting each such subtree to its terminal root reduces $P_0$ to the tree $T_0$, so $P_0$ is connected and acyclic, and thus a tree containing $U$. Counting each shared edge once gives $w(P_0)\le w(T_0)\le 2\OPT_Q(U)$ by Lemma~\ref{lem:metric-mst}. To construct $P_0$ in $O(N)$ time, we walk up the owner forest from both endpoints of each of the $|U|-1$ cross-region edges of $T_0$, marking visited vertices and stopping each walk upon reaching a terminal root or an already marked vertex so that each of the at most $N$ owner-forest edges is traversed at most once.
\end{proof}

\paragraph{MST under graph augmentation.}
During an update phase, our algorithms augment the frozen terminal metric closure $K_S(R)$ with a small set of edges. Rather than maintaining the dense graph $K_S(R)\cup C$, the following general MST sparsification lemma shows that replacing $K_S(R)$ by its precomputed MST $T_0$ preserves the minimum spanning tree weight exactly.

\begin{lemma}[MST under graph augmentation]\label{lem:mst-augmentation}
Let $K$ be a connected weighted graph, and let $T_0$ be an MST of $K$. Augment $K$ by adding a set $C$ of weighted edges, either connecting existing vertices of $K$ or introducing new vertices, such that $K\cup C$ is connected. Then
\[
\mst(K\cup C)=\mst(T_0\cup C).
\]
Parallel edges are allowed and retained as distinct edges in $K\cup C$ and $T_0\cup C$.
\end{lemma}
\begin{proof}
Since $T_0\cup C\subseteq K\cup C$, we have $\mst(K\cup C)\le\mst(T_0\cup C)$. Conversely, let $T$ be an MST of $K\cup C$. For any edge $e=uv\in E(T)\cap E(K\setminus T_0)$, the cycle property of the MST $T_0$ of $K$ ensures that every edge on the unique $u$--$v$ path in $T_0$ has weight at most $w(e)$. Removing $e$ splits $T$ into two components, and at least one edge $f$ on the $u$--$v$ path in $T_0$ crosses this cut; replacing $e$ by $f$ yields a spanning tree of $K\cup C$ with weight at most $w(T)$ and one fewer edge from $K\setminus T_0$. Repeating this exchange eliminates all edges of $K\setminus T_0$, producing a spanning tree of $T_0\cup C$ of weight at most $w(T)=\mst(K\cup C)$.
\end{proof}

\section{Insertion-only Algorithm}
\label{sec:insertion-only}
\label{inc:algorithm}
We introduce our framework in a simpler form for insertion-only updates. Our approach is to divide the update sequence into phases of $B$ insertions and combine two components that together help us avoid recomputing terminal distances:
\begin{enumerate}[nosep,label=(\roman*)]
  \item \emph{Frozen structures}, computed once at the beginning of each phase on top of the dynamic spanner $S_t$, provide graph-wide routing information, including a baseline terminal-metric MST $T_0$, nearest-terminal owners $a(x)\in R$, and approximate all-pairs distances in $S$.
  \item \emph{Steiner augmentation structures}, updated after each insertion on the at most $2B$ endpoints touched by the current phase, capture information associated with newly inserted edges, such as routes that use new edges and certain augmentations of $T_0$.
\end{enumerate}
Because no edges are deleted in the insertion-only model, all phase-start routes in $S$ remain valid throughout the phase; new insertions can only create cheaper connections between terminal owner regions, which we incorporate into $T_0$ via the MST augmentation property (Lemma~\ref{lem:mst-augmentation}). We describe the algorithm below and defer full implementation details and proofs to Appendix~\ref{app:insertion-only}.

\paragraph{Frozen structures.}
In the background, we maintain the dynamic spanner $S_t\subseteq G_t$ of Theorem~\ref{thm:dynamic-spanner} with stretch $\sigma=(2s-1)(1+\delta)$ (for a fixed integer $s\ge 2$ and constant $0<\delta<1$) and expected size $\Oh(sn^{1+1/s})$. At the beginning of each phase, we copy the current spanner into a frozen snapshot $S$ and compute two static structures on $S$:
\begin{enumerate}[nosep,label=(\arabic*)]
  \item We run Mehlhorn's construction (Theorem~\ref{thm:mehlhorn}) on $(S,R)$ to obtain a terminal-metric MST $T_0$ of $K_S(R)$, each vertex's nearest terminal owner $a(x)\in R$, its owner distance $\delta(x)=d_S(a(x),x)$, and the rooted owner forest in $S$.
  \item We build the Thorup--Zwick path-reporting distance oracle (Theorem~\ref{inc:oracle}) on $S$ with stretch $A=2r-1$ (for a fixed integer $r\ge 2$), which answers queries with distance estimates $d_S(x,y)\le \widehat d_S(x,y)\le A d_S(x,y)$ and realizing walks in $S$.
\end{enumerate}

\paragraph{Steiner augmentation structures.}
Let $F$ denote the set of edges inserted so far in the current phase and let $X$ be the set of their endpoints, so $|F|\le B$ and $|X|\le 2B$. Because any route in $S \cup F$ that improves upon $S$ must traverse at least one edge of $F$, we maintain a weighted multigraph $L_F$ on the endpoint set $X$ with two types of edges:
\begin{itemize}[nosep]
  \item every newly inserted edge $e\in F$ with its actual weight $w(e)$; and
  \item for every pair of distinct vertices $x,y\in X$, a virtual edge of weight $\widehat d_S(x,y)$ queried from the phase-start distance oracle on $S$.
\end{itemize}
As edges are inserted into $F$, we maintain exact all-pairs shortest-path distances $D_F(x,y)=d_{L_F}(x,y)$ and a next-hop matrix on $L_F$ in $O(|X|^2)=O(B^2)$ time per insertion. Thus, $D_F(x,y)$ records the cheapest walk between $x$ and $y$ that combines newly inserted edges in $F$ with oracle paths in $S$.

\paragraph{Maintaining the augmented terminal MST.}
To translate shortest paths in $L_F$ into connections between terminals in $R$, we route each endpoint $x\in X$ to its frozen terminal owner $a(x)\in R$ along the owner forest in $S$. Specifically, after updating $D_F$ on an insertion, for every pair $x,y\in X$ with $a(x)\neq a(y)$, we create or update a candidate edge between terminals $a(x)$ and $a(y)$ of weight
\begin{equation}\label{inc:candidate}
c_{xy}=\delta(x)+D_F(x,y)+\delta(y),
\end{equation}
which never increases during the phase because $\delta(\cdot)$ is fixed and insertions can only decrease $D_F(x,y)$. Each candidate edge $a(x)a(y)$ represents a walk in the current graph formed by concatenating the owner-forest path from $a(x)$ to $x$, the shortest path in $L_F$ from $x$ to $y$ (with virtual edges expanded via the path-reporting oracle on $S$), and the owner-forest path from $y$ to $a(y)$. Letting $C_F$ denote the set of these $O(B^2)$ candidate edges on $R$, we maintain a minimum spanning tree $\mathcal M$ of the augmented graph $T_0\cup C_F$. By the MST augmentation property,
\begin{equation}\label{inc:certificate}
\mst(T_0\cup C_F)=\mst(K_S(R)\cup C_F),
\end{equation}
so maintaining an MST of $T_0\cup C_F$ achieves the exact same weight as maintaining an MST of the full terminal metric $K_S(R)$ augmented with $C_F$. Since candidate weights only decrease during a phase, we can maintain $\mathcal M$ dynamically in $O(\log n)$ time per candidate update using a link--cut tree~\cite{ST83}, taking $O(B^2\log n)$ time per insertion. When an explicit Steiner tree is requested, we coherently expand the owner-forest routes of $\mathcal M$ in $S$, extract a spanning forest of the $L_F$-paths for the $O(B)$ selected candidate edges and materialize its virtual edges via the path-reporting oracle, and finally delete cycles and prune non-terminal leaves.

After $B$ insertions occur, we start a new phase and rebuild the frozen structures on the current spanner. We also reset $F$, $X$, and $C_F$. Balancing this periodic rebuild cost against the per-insertion update cost yields our amortized update time. In particular, choosing $B=\left\lceil n^{(1+1/s+1/r)/3}\right\rceil$ gives the guarantee stated in Theorem~\ref{thm:main-incremental}.

The fully dynamic algorithm in Section~\ref{sec:fd-design} retains the same framework, combining \emph{frozen structures} with within-phase \emph{Steiner augmentation structures} and mapping endpoints of newly inserted edges to their frozen terminal owners $a(u),a(v)\in R$. However, deleting an edge may invalidate many routes represented by the terminal MST $T_0$. To \textit{localize} this impact, we will expand $T_0$ into a tree $P_0$, where a deletion removes only the corresponding edge, and represent both newly inserted edges (via their owners $a(u),a(v)$) and deleted edges of $P_0$ (via their endpoints $u,v$) by zero-weight spokes to an artificial hub $h$. Rather than maintaining the auxiliary multigraph $L_F$ and candidate edges $C_F$, a top-tree dynamic program will select portions of $P_0$ to retain. A dynamic distance sparsifier estimates the reconnection cost, while Mehlhorn's construction on the current spanner reconnects these portions when an explicit solution is requested.
\section{Fully Dynamic Algorithm}
\label{sec:fd-design}
This section is devoted to our main result, a fully dynamic algorithm for the Steiner tree problem that simultaneously achieves sublinear-in-$n$ update time and a constant factor approximation. Our algorithm builds on the dynamic spanner of Chuzhoy and Parter~\cite{CP25} (cf.\ Theorem~\ref{thm:cp-spanner}) and the HLS distance sparsifier (cf.\ Theorem~\ref{thm:hls}) to maintain a sparse tree spanning the terminals, while dynamically reconnecting its surviving pieces as edges are inserted and deleted. We present the overall algorithm in Section~\ref{sec:fd-algorithm}, followed by the details of its two core components: the maintenance of a minimum Steiner tree on a certain augmented tree in Section~\ref{sec:optimizer}, and the component repair procedure in Section~\ref{sec:repair}. We then establish the approximation guarantee in Section~\ref{sec:fd-approx} and conclude with the update-time, reporting, and space analysis in Section~\ref{sec:fd-runtime}.

\subsection{Algorithm}
\label{sec:fd-algorithm}
We divide the update sequence into phases of $B$ updates each. Let $G^\circ$ denote the graph at the start of a phase; objects described as \emph{frozen} are computed from $G^\circ$ at the phase boundary and remain fixed throughout the phase, whereas the current graph $G_t$ continues to evolve. Across all phases, our algorithm maintains two persistent data structures on the current graph $G_t$: CP's deterministic dynamic $\sigma$-spanner $S_t\subseteq G_t$ with $\sigma=\sigma_\nu$ and an explicit HLS distance sparsifier $H_t$ on a designated vertex set $U_t\subseteq V$ containing a fixed root terminal $r_0\in R$. To respect the integer-weight requirements of Theorems~\ref{thm:cp-spanner} and~\ref{thm:hls}, all internal structures operate on the rounded integer weights $\widehat w(e)=\lceil w(e)/\varepsilon\rceil$ from~\eqref{eq:rounding}; we write $w$ for $\widehat w$ through Theorem~\ref{thm:fd-phase-approx} to ease notation. We now describe how the algorithm builds its frozen structures at the start of a phase, maintains its Steiner augmentation structures, and reports an explicit Steiner tree upon query.

\paragraph{Frozen structures.}
At the beginning of each phase, we take a frozen copy $S\subseteq G^\circ$ of the current dynamic spanner $S_t$ and construct the following structures in two steps:
\begin{enumerate}[label=(\arabic*)]
\item \emph{Terminal-metric MST and owner forest.} We run Mehlhorn's construction (Theorem~\ref{thm:mehlhorn}) on $S$ with terminal set $R$ to obtain each vertex's nearest terminal owner $a(x)\in R$, its owner distance $\delta(x)=d_S(a(x),x)$, the rooted owner forest, and an MST $T_0$ of $K_S(R)$.
\item \emph{Coherent tree expansion.} Because a single edge deletion could invalidate many routes of $T_0$ simultaneously, we apply Lemma~\ref{lem:coherent-expansion} to expand $T_0$ along the owner forest into a single tree $P_0\subseteq S$ spanning $R$, retaining each shared  edge only once.
\end{enumerate}

\paragraph{Steiner augmentation structures.}
To track updates during a phase and decide which surviving pieces of $P_0$ to retain, we augment $P_0$ at the start of the phase with an artificial vertex $h$, called \emph{hub}, and a zero-weight artificial edge $zh$, called a \emph{spoke}, for every vertex $z\in V(P_0)$. Each edge of $P_0$ and each spoke may be enabled or disabled; we denote by $J_t$ the subgraph of enabled edges at time $t$, and call a vertex $z$ whose spoke $zh$ is enabled an \emph{active vertex} (or an \emph{active terminal} when $z\in R$). At the start of a phase, all edges of $P_0$ are enabled and all spokes are disabled. As updates arrive during the phase, we update $J_t$ as follows:
\begin{itemize}
\item \emph{Insertion handling.} Inserting an edge $uv$ enables the spokes $a(u)h$ and $a(v)h$ at the endpoints' \emph{frozen terminal owners} $a(u),a(v)\in R$, providing a  zero-cost connection through $h$ that allows the algorithm to omit expensive old routes in $P_0$.
\item \emph{Deletion handling.} Deleting an edge $uv\in E(P_0)$ disables $uv$ in $P_0$ and enables the spokes $uh$ and $vh$ at its \emph{endpoints} $u,v\in V(P_0)$ (which need not be terminals), providing a zero-cost bridge across the resulting gap in $P_0$. Deleting an edge outside $E(P_0)$ leaves $J_t$ unchanged.
\end{itemize}
Once enabled, a spoke remains enabled until the phase ends. Letting $A_t\subseteq R$ denote the set of \emph{active terminals}, namely the frozen owners of all endpoints of edges inserted during the phase, and $D_t\subseteq E(P_0)$ denote the edges of $P_0$ deleted during the phase, the set of active vertices at time $t$ is
\begin{equation}\label{eq:fd-ports}
Z_t=A_t\cup\operatorname{end}(D_t),
\end{equation}
where $\operatorname{end}(D_t)$ consists of the endpoints of all edges in $D_t$; the resulting auxiliary graph is
\begin{equation}\label{eq:fd-hub-graph}
J_t=(P_0-D_t)\ \cup\ \{zh:z\in Z_t\},
\qquad w(zh)=0.
\end{equation}
Note that sets $A_t$, $D_t$, and $Z_t\subseteq V(P_0)$ only grow within a phase and each has size $O(B)$.

\paragraph{Reconnecting surviving subtrees.}
After updating $J_t$ and forwarding the graph update to $S_t$ and $H_t$, our algorithm maintains its implicit Steiner tree in three steps:
\begin{enumerate}[label=(\arabic*)]
\item \emph{Maintaining an exact Steiner tree on $J_t$ (Section~\ref{sec:optimizer}).} Using dynamic programming over a fixed top-tree hierarchy on $P_0$, we maintain a minimum Steiner tree $T_t^J$ connecting $R$ in $J_t$ that lexicographically minimizes
\[
\bigl(\text{weight},\ \text{number of spokes},\ \text{number of edges}\bigr).
\]
Removing the artificial vertex $h$ and its selected spokes from $T_t^J$ leaves a surviving forest $F_t=T_t^J-h\subseteq P_0-D_t$. 
When $T_t^J$ uses $h$, lexicographic optimality ensures that each component of $F_t$ contains a terminal in $R$ and a unique selected spoke to $h$ (Lemma~\ref{opt:representatives}). Collecting the endpoints in $P_0$ of these selected spokes gives a set $Q_t\subseteq Z_t\cup\{r_0\}$ of component \emph{representatives} of size $O(B)$ (with $Q_t=\{r_0\}$ when $h$ is unused).
\item \emph{Estimating reconnection cost via distance sparsifier (Section~\ref{sec:repair}).} Because each component of $F_t$ connects its terminals to its representative in $Q_t$, reconnecting $Q_t$ in $G_t$ suffices to connect $R$. Since $Q_t$ may change substantially after a single update, we maintain the HLS distance sparsifier $H_t$ on the monotonically growing superset $U_t=Z_t\cup\{r_0\}$, inserting the $O(1)$ newly active vertices per update and resetting $U_t$ to $\{r_0\}$ at phase boundaries. We run Mehlhorn's construction (Theorem~\ref{thm:mehlhorn}) on $H_t$ with terminal set $Q_t$ to compute an MST $M_H$ of $K_{H_t}(Q_t)$, whose scaled weight $\sigma\,w_{K_{H_t}(Q_t)}(M_H)$ upper-bounds the cost of reconnecting $Q_t$ in $S_t$.
\item \emph{Implicit representation and tree reporting.} The maintained implicit solution consists of the dynamic-programming backpointers on the top tree (encoding $F_t$ and $Q_t$) together with the scalar upper-bound cost estimate $c_t=w(F_t)+\sigma\,w_{K_{H_t}(Q_t)}(M_H)$ combining the top-tree root weight $w(F_t)$ with the scaled sparsifier MST weight. When an explicit Steiner tree is requested, we follow the backpointers to recover $F_t$, run Mehlhorn's construction (Theorem~\ref{thm:mehlhorn}) and coherent tree expansion (Lemma~\ref{lem:coherent-expansion}) on the current spanner $S_t$ with terminal set $Q_t$ to obtain a tree $A_S\subseteq S_t$ connecting $Q_t$, and take the union $F_t\cup A_S$, removing cycles and pruning nonterminal leaves.
\end{enumerate}

Algorithm~\ref{alg:fd-initialize} summarizes initialization, reconstruction, update, and materialization procedures.

\begin{figure}[tb]
\begin{pseudocode}{Fully Dynamic Steiner Tree Algorithm}
\label{alg:fd-initialize}
\label{alg:fd-materialize}
\label{sec:pseudocode}
\begin{algorithmsteps}
\item \textbf{procedure} \textsc{Initialize}$(G_0,R)$
\item \hspace*{1em} Set rounded weights $\widehat w(e)\gets\lceil w(e)/\varepsilon\rceil$ for all $e\in E_0$. Fix a root terminal $r_0\in R$ and initialize the deterministic dynamic $\sigma$-spanner $S_t$ (Theorem~\ref{thm:cp-spanner}) and HLS sparsifier $H_t$ with designated set $U_t\gets\{r_0\}$ (Theorem~\ref{thm:hls}) on $G_0$ (under $\widehat w$).
\item \hspace*{1em} Call \textsc{StartPhase}().

\bigskip
\item \textbf{procedure} \textsc{StartPhase}() \hfill \emph{$\triangleright$ invoked at initialization and every $B$ updates}
\item \hspace*{1em} Delete each $z\in U_t\setminus\{r_0\}$ from the HLS sparsifier $H_t$ to reset the designated set to $U_t\gets\{r_0\}$, and set the active-vertex set to $Z_t\gets\varnothing$.
\item \hspace*{1em} Copy $S_t$ into a frozen snapshot $S\subseteq G^\circ$ and run Mehlhorn's construction (Theorem~\ref{thm:mehlhorn}) on $(S,R)$ to obtain terminal owners $a(x)$, the owner forest, and an MST $T_0$ of $K_S(R)$.
\item \hspace*{1em} Expand $T_0$ along the owner forest into the tree $P_0\subseteq S$ (Lemma~\ref{lem:coherent-expansion}).
\item \hspace*{1em} Initialize the top-tree data structure (Section~\ref{sec:optimizer}) on $J_t=P_0$ under $\widehat w$ (all $P_0$ edges enabled, all spokes disabled), and set $Q_t\gets\{r_0\}$ and $c_t\gets\widehat w(P_0)$.

\bigskip
\item \textbf{procedure} \textsc{Update}$(\mathrm{op}, uv, w)$ \hfill \emph{$\triangleright$ see Remark~\ref{rem:weight-changes} for weight changes}
\item \hspace*{1em} If $\mathrm{op}=\text{insert}$, set $\widehat w(uv)\gets\lceil w/\varepsilon\rceil$. Forward the update to $G_t$, $S_t$, and $H_t$.
\item \hspace*{1em} \textbf{If} $uv$ is inserted: enable spokes $a(u)h, a(v)h$ in $J_t$; for each $z\in\{a(u),a(v)\}\setminus U_t$, insert $z$ into the designated set $U_t$ of $H_t$; and add $a(u),a(v)$ to $Z_t$.
\item \hspace*{1em} \textbf{If} $uv\in E(P_0)$ is deleted: disable $uv$ and enable spokes $uh,vh$ in $J_t$; for each $z\in\{u,v\}\setminus U_t$, insert $z$ into the designated set $U_t$ of $H_t$; and add $u,v$ to $Z_t$.
\item \hspace*{1em} \emph{Select surviving forest (Section~\ref{sec:optimizer}):} Update the top-tree DP on $J_t$ to maintain the optimal tree $T_t^J$ and extract the representatives $Q_t$ of $F_t=T_t^J-h$.
\item \hspace*{1em} \emph{Reconnect components (Section~\ref{sec:repair}):} Run Mehlhorn's construction on $(H_t,Q_t)$ to compute an MST $M_H$ of $K_{H_t}(Q_t)$ and record $c_t\gets\widehat w(F_t)+\sigma\,w_{K_{H_t}(Q_t)}(M_H)$ (cost queries return $\varepsilon c_t$).

\bigskip
\item \textbf{procedure} \textsc{Materialize}()
\item \hspace*{1em} Recover the forest $F_t\subseteq P_0$ by following the top-tree DP backpointers.
\item \hspace*{1em} Run Mehlhorn's construction (Theorem~\ref{thm:mehlhorn}) on $(S_t,Q_t)$ to compute an MST $M_S$ of $K_{S_t}(Q_t)$, and expand $M_S$ along its owner forest in $S_t$ into a tree $A_S\subseteq S_t$ (Lemma~\ref{lem:coherent-expansion}).
\item \hspace*{1em} In $F_t\cup A_S$, delete cycles and prune nonterminal leaves to return a Steiner tree for $R$.
\end{algorithmsteps}
\end{pseudocode}
\end{figure}

\subsection{Minimum Steiner Tree Maintenance on Hub-Augmented Tree}\label{sec:optimizer}
In this section, we describe how to dynamically maintain a
\emph{minimum} Steiner tree $T_t^J$ connecting $R$ in $J_t=(P_0-D_t)\cup\{zh:z\in Z_t\}$. The graph $J_t$ has a special tree-plus-star structure: all of its  non-spoke edges belong to the fixed tree $P_0$, and every other edge is a zero-weight spoke incident to the hub $h$. This structure helps us maintain a Steiner tree of minimum cost in $O(\log n)$ time per update via dynamic programming. 
We first give a static algorithm for finding a minimum Steiner tree in $J_t$, and then explain how to maintain this optimum dynamically.

Among all Steiner trees connecting $R$ in $J_t$, we minimize the lexicographic triple
\begin{equation}\label{eq:opt-triple}
  \Lambda(Y)=\bigl(w(Y),\;\text{number of selected spokes},\;\text{number of selected edges}\bigr).
\end{equation}
The second coordinate rules out unnecessary free hub spokes, while the third prefers fewer edges. Removing the hub $h$ and its incident spokes from $T_t^J$ yields the surviving forest
\[
F_t=T_t^J-h.
\]
If $h$ is used in $T_t^J$, we define the representative set $Q_t$ to consist of the endpoints in $P_0$ of the selected spokes; if $h$ is unused, we set $Q_t=\{r_0\}$. Minimizing~\eqref{eq:opt-triple} guarantees that each component of $F_t$ spans at least one terminal and has a unique representative:

\begin{lemma}[One representative per retained component]\label{opt:representatives}
    Every component of $F_t$ contains at least one original terminal and exactly one representative. If the hub is used, $F_t$ has at least two components, each connected to the hub by exactly one selected spoke. If the hub is unused, $F_t$ is connected and $Q_t=\{r_0\}$.
\end{lemma}
\begin{proof}
    Every connected component of $F_t$ must contain a terminal in $R$: otherwise, pruning that component and its incident spokes from the tree $T_t^J$ would preserve connectivity on $R$ while strictly improving $\Lambda(T_t^J)$ (decreasing either weight or spoke count), contradicting optimality. If $F_t$ has multiple components, each must be incident to at least one spoke in $T_t^J$ to reach the others through $h$, and at most one spoke because two spokes into the same component of $F_t$ would create a cycle through $h$. Finally, if $F_t$ has a single component, it already connects $R$, so selecting any spoke would increase the spoke count without reducing weight; hence $h$ is unused and $Q_t=\{r_0\}$.
\end{proof}

In the remainder of this subsection, we show how to compute and maintain $T_t^J$. We first decompose $P_0$ into a hierarchy of connected subtrees of height $O(\log n)$ using a top tree~\cite{AHLT05}, and assign each edge of $P_0$ and hub spoke of $J_t$ to a unique cluster where the dynamic program chooses whether to include it. Because each cluster interacts with the rest of $J_t$ through at most two boundary vertices in $P_0$ and the hub $h$, we can summarize all partial solutions within a cluster using a constant-size DP table of $22$ states. Finally, we show that each edge or spoke update only recomputes the $O(\log n)$ DP tables along a single cluster-to-root path, and that following backpointers from the root extracts the representatives $Q_t$ in $O(|Q_t|\log n)$ time.

\paragraph{A fixed top-tree hierarchy on $P_0$.}
To maintain $T_t^J$ under edge and spoke changes, we organize the fixed tree $P_0$ using a top tree~\cite{AHLT05}. This is a balanced binary hierarchy that recursively merges adjacent connected edge sets of $P_0$ into larger pieces. A node of the hierarchy represents a \emph{cluster}: a connected set of edges of $P_0$ and their endpoints. The leaves represent individual edges of $P_0$. Two sibling clusters have disjoint edge sets and share exactly one vertex.

To describe how a cluster connects to the rest of $P_0$, we identify its \emph{boundary vertices}. A vertex of a cluster $C$ is a \emph{boundary vertex} (written $v\in\partial C$) if it is incident to an edge of $P_0\setminus C$ or equals the fixed root terminal $r_0\in R$, which we designate as an \emph{external boundary vertex}~\cite{AHLT05}. Keeping $r_0$ in $\partial C$ for every cluster containing it (even when no edge of $P_0\setminus C$ is incident to $r_0$) ensures that at the root cluster we have $\partial P_0=\{r_0\}$ and all terminals in $R$ are connected to $r_0$.

\begin{figure}[tb]
\centering
\begin{tikzpicture}[font=\small,bag/.style={draw,rounded corners=2pt,align=center,minimum width=2.5cm,minimum height=0.78cm,fill=blue!3},hlink/.style={draw,gray!70},vtx/.style={circle,fill=black,inner sep=1.7pt}]
\node[bag] (root) at (0,3.1) {$e_1,e_2,e_3,e_4$\\$\partial=\{r_0\}$};
\node[bag] (left) at (-3,1.65) {$e_1,e_2$\\$\partial=\{r_0,b\}$};
\node[bag] (right) at (3,1.65) {$e_3,e_4$\\$\partial=\{b\}$};
\node[bag] (l1) at (-4.5,0.2) {$e_1$\\$\partial=\{r_0,a\}$};
\node[bag] (l2) at (-1.5,0.2) {$e_2$\\$\partial=\{a,b\}$};
\node[bag] (l3) at (1.5,0.2) {$e_3$\\$\partial=\{b,c\}$};
\node[bag] (l4) at (4.5,0.2) {$e_4$\\$\partial=\{c\}$};
\draw[hlink] (root)--(left) (root)--(right) (left)--(l1) (left)--(l2) (right)--(l3) (right)--(l4);
\node[vtx,label=below:$r_0$] (p0) at (-6,-1.35) {};
\node[vtx,label=below:$a$] (p1) at (-3,-1.35) {};
\node[vtx,label=below:$b$] (p2) at (0,-1.35) {};
\node[vtx,label=below:$c$] (p3) at (3,-1.35) {};
\node[vtx,label=below:$d$] (p4) at (6,-1.35) {};
\draw[thick] (p0)--node[above]{$e_1$}(p1)--node[above]{$e_2$}(p2)--node[above]{$e_3$}(p3)--node[above]{$e_4$}(p4);
\end{tikzpicture}
\caption{An example top-tree hierarchy on the four-edge path $P_0$ shown below it. Each box lists its edges and boundary set. The exposed terminal $r_0$ remains a boundary at every containing cluster. The vertex $b$ becomes internal only at the root merge; $d$ is already internal at its singleton leaf. The hub and spokes are absent from this hierarchy.}
\label{fig:cluster-hierarchy}
\end{figure}
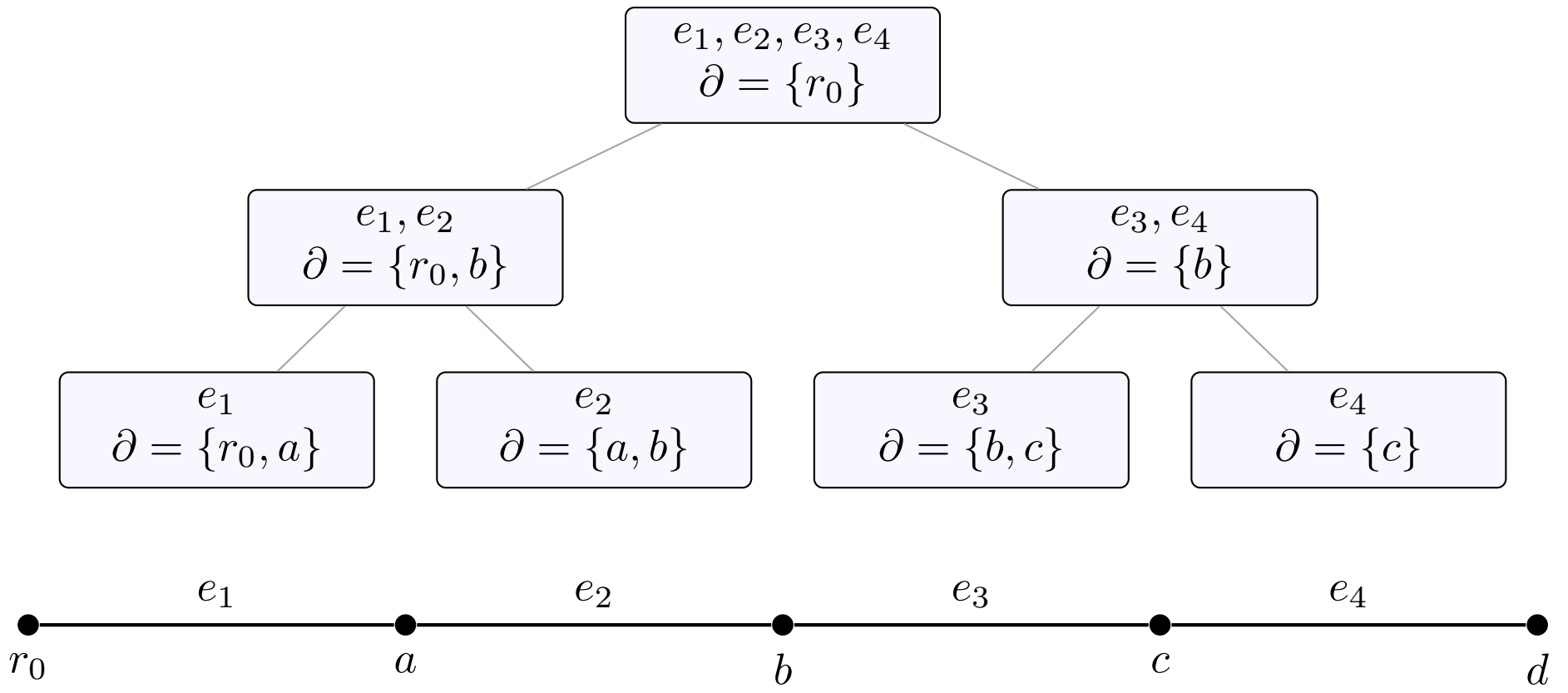

By~\cite{AHLT05}, building the hierarchy on the $O(n)$ edges of $P_0$ and exposing $r_0$ takes $O(n\log n)$ time and produces $O(n)$ clusters of height $O(\log n)$ such that every cluster $C$ has $|\partial C|\le 2$, as illustrated in Figure~\ref{fig:cluster-hierarchy}. Crucially, we build this hierarchy on all edges of $P_0$ (including disabled ones), and the hub spokes are \emph{not} part of the hierarchy. Whether an edge or spoke is enabled determines whether it can be selected by our dynamic program, without altering the shape or boundary sets of the clusters; thus, the cluster tree itself remains fixed throughout the phase, and only the clusters' DP entries change upon updates.

\paragraph{Assigning each edge and spoke to a unique cluster.}
As our bottom-up dynamic program moves up the hierarchy, it builds partial Steiner trees by choosing whether to include or omit each enabled edge of $J_t$. For edges of $P_0$, this choice happens at the leaves. Because the hub spokes are not part of the top tree on $P_0$, however, we must specify at which cluster the dynamic program chooses whether to include each spoke $vh$, so that every spoke is considered exactly once before its  endpoint $v \in V(P_0)$ disappears from the cluster boundary. We have
\begin{itemize}[nosep]
  \item Each edge $e\in E(P_0)$ is chosen (included or omitted) at its singleton leaf cluster $\{e\}$.
  \item Each potential spoke $vh$ with $v\neq r_0$ is chosen at the unique cluster where $v$ first stops being a boundary vertex, i.e., the smallest cluster containing all edges of $P_0$ incident to $v$. Once all edges of $P_0$ incident to $v$ lie inside a cluster, $v$ will never appear on the boundary of any ancestor cluster. This cluster is either a singleton leaf if $\deg_{P_0}(v)=1$ (such as $d$ in Figure~\ref{fig:cluster-hierarchy}) or the merge of two child clusters $A$ and $B$ sharing $v$ once no edge incident to $v$ remains outside $A\cup B$. The root spoke $r_0h$ is chosen at the root cluster $P_0$.
\end{itemize}
During a merge of two sibling clusters, only their single shared vertex can stop being a boundary vertex, so each merge introduces at most one spoke choice (plus $r_0h$ at the root). For each edge of $P_0$ and spoke, we store a pointer to the cluster where its choice is made, allowing that cluster's DP table to be located in $O(1)$ time when the edge or spoke is enabled or disabled.

\paragraph{Cluster states and dynamic-programming transitions.}
To build a Steiner tree connecting $R$ to $r_0$, the dynamic program maintains partial solutions at each cluster $C$ by selecting enabled edges of the cluster and enabled spokes whose decisions are made in that cluster or its descendants. Once all choices inside $C$ have been made, vertices inside $C$ can connect to the rest of $J_t$ outside $C$ only through the boundary vertices $\partial C$ (via edges of $P_0$ outside $C$) or through the hub $h$ (via spokes outside $C$). We therefore call
\[
I_C=\partial C\cup\{h\},\qquad |I_C|\le 3,
\]
the \emph{interfaces} of $C$. For a partial solution inside $C$ to be extendable to a valid Steiner tree of $J_t$, it must satisfy two basic conditions:
\begin{enumerate}[nosep]
  \item \emph{Connectivity:} Every terminal in $C$ must be connected to at least one interface in $I_C$. Otherwise, since no edge outside $C$ can ever touch an internal vertex of $C$, that terminal would be permanently trapped inside $C$ with no way to reach $r_0$. (Because $r_0$ remains in $\partial C$ for every cluster containing it, any terminal already connected to $r_0$ automatically satisfies this condition.)
  \item \emph{Acyclicity:} The edges and spokes selected within $C$ must contain no cycle.
\end{enumerate}
A feasible partial solution inside $C$ interacts with the rest of $J_t$ as follows. If a component inside $C$ contains a terminal, edges outside $C$ must eventually connect it to all other terminal-bearing components, and they can reach it through \emph{any} interface in $I_C$ belonging to that component; conversely, if two interfaces in $I_C$ already lie in the same component inside $C$, connecting them again outside $C$ would create a cycle.

We therefore encode a partial solution in $C$ by partitioning its interfaces $I_C$ into connected components within $C$ and assigning each group a terminal bit $b_i\in\{0,1\}$ equal to $1$ if that component in $C$ contains a terminal in $R$, and $0$ otherwise. For instance, when $\partial C=\{u,v\}$, the states
\[
\{u\}_1\mid\{v\}_1\mid\{h\}_0
\qquad\text{and}\qquad
\{u,v\}_1\mid\{h\}_0
\]
express different connectivity needs: in the first state, $u$ and $v$ lie in separate terminal-bearing components inside $C$, so edges outside $C$ must eventually connect both $u$ and $v$; in the second state, $u$ and $v$ are already joined inside $C$, so connecting through either $u$ or $v$ suffices to reach all terminals in their shared component. With up to three interfaces $u,v,h$, there are at most $22$ states:

\begin{center}
\begin{tabular}{lcc}
\toprule
Grouping & Encodings & Number of states \\
\midrule
Three separate groups
& $\{u\}_{b_1}\mid\{v\}_{b_2}\mid\{h\}_{b_3}$
& $2^3=8$ \\
\addlinespace
Two groups
& $\begin{gathered}
\{u,v\}_{b_1}\mid\{h\}_{b_2}\\
\{u,h\}_{b_1}\mid\{v\}_{b_2}\\
\{v,h\}_{b_1}\mid\{u\}_{b_2}
\end{gathered}$
& $3\cdot 2^2=12$ \\
\addlinespace
One group
& $\{u,v,h\}_{b_1}$
& $2$ \\
\bottomrule
\end{tabular}
\end{center}
Each cluster's DP table stores, for each state, the minimum cost triple~\eqref{eq:opt-triple} over all feasible partial solutions inducing that state (or $+\infty$ if none exists), together with backpointers to the local edge/spoke choices and child states achieving it. We now describe how to compute these DP tables bottom-up in three stages: initializing singleton leaves, merging sibling clusters, and selecting the final tree at the root.
\begin{itemize}[nosep]
  \item \emph{Singleton leaves.} A leaf cluster $C=\{e\}$ with $e=uv$ has at most two edges assigned to it ($e$ and, for any endpoint $v\notin\partial C$, the spoke $vh$), giving at most four enabled subsets (or at most eight if the singleton is also the root and decides $r_0h$). We reject any subset that leaves a terminal endpoint of $e$ with no interface in $I_C$. For each feasible subset, we record its induced partition and terminal bits on $I_C$ with cost $(w(e),0,1)$ if $e$ is selected (and $(0,0,0)$ otherwise), plus $(0,1,1)$ per selected spoke, keeping the minimum cost and local choices for each state in $O(1)$ time.
  \item \emph{Internal merges.} To compute the table of $C=A\cup B$ from its children $A$ and $B$ sharing boundary vertex $v$, we enumerate the at most $22\times 22\times 2$ combinations of finite-cost child states $(s_A,s_B)$ and the local spoke choice at $v$ if $v\notin\partial C$ (or $\times 4$ at the root to also choose $r_0h$). Using only the recorded interface groups and terminal bits of $s_A$ and $s_B$ (without examining internal vertices or paths), we unite the groups connected through $v$, $h$, or $vh$, OR their terminal bits, and apply two feasibility checks:
  \begin{enumerate}[nosep,label=(\roman*)]
    \item \emph{Acyclicity:} Because $A$ and $B$ share no edges of $P_0$ or previously chosen spokes and overlap only at $v$ and $h$, a cycle arises if and only if $v$ and $h$ already belong to the same group in both $s_A$ and $s_B$ (e.g., $\{u,v,h\}_1$ and $\{v,w,h\}_0$, where each child provides a distinct $v$--$h$ path), or belong to the same group in one child while the spoke $vh$ is also selected (or, at the root, $r_0$ and $h$ already lie in the same merged group when $r_0h$ is selected). Any such combination is rejected.
    \item \emph{No stranded terminals:} If $v$ becomes internal ($v\notin\partial C$), its merged group cannot be reached by any edges outside $C$ and must therefore contain at least one surviving interface in $I_C=\partial C\cup\{h\}$ whenever its merged terminal bit is~$1$ (rejecting, e.g., $\{u\}_0\mid\{v\}_1\mid\{h\}_0$ and $\{v\}_0\mid\{w\}_0\mid\{h\}_0$ when $v\notin\partial C$ and $vh$ is omitted, since $\{v\}_1$ loses its last interface).
  \end{enumerate}
  For every combination passing both checks, we project its partition onto $I_C$ by dropping $v$ if $v\notin\partial C$ (for example, merging $\{u\}_0\mid\{v,h\}_1$ and $\{v,w\}_0\mid\{h\}_0$ with $v\notin\partial C$ and $vh$ omitted yields $\{u\}_0\mid\{w,h\}_1$) and update that state's table entry in $O(1)$ time with the sum of the two child cost triples plus $(0,1,1)$ per selected local spoke.
  \item \emph{Root selection.} At the root $P_0$, where $I_{P_0}=\{r_0,h\}$, we accept the states in which there is a single terminal-bearing group and it contains $r_0$---namely $\{r_0,h\}_1$ (when the hub is used) and $\{r_0\}_1\mid\{h\}_0$ (when the hub is unused), whereas $\{r_0\}_1\mid\{h\}_1$ is rejected because its terminals lie in two disconnected components---and take the minimum cost triple over the accepted states.
\end{itemize}

\begin{lemma}[Static top-tree DP on $J_t$]\label{opt:exact}
Given the top-tree hierarchy on $P_0$, the DP tables of all clusters can be computed in $O(n)$ time and $O(n)$ space, and the root table returns a Steiner tree $T_t^J$ connecting $R$ in $J_t$ that minimizes~\eqref{eq:opt-triple}.
\end{lemma}
\begin{proof}
The leaf and merge rules reject cycles and stranded terminals, and the root check requires all terminals to be connected to $r_0$, ensuring feasibility. Optimality follows by induction on the cluster tree: any two partial solutions within a cluster $C$ that induce the same partition and terminal bits on $I_C$ are interchangeable with respect to any completion outside $C$, and the transitions exhaust all local edge and spoke choices. Since there are $O(n)$ clusters and each table has $O(1)$ states updated in $O(1)$ time, total time and space are $O(n)$.
\end{proof} 

\paragraph{Dynamic maintenance and representative extraction.}
Because the top-tree hierarchy on $P_0$ has height $O(\log n)$ and remains fixed throughout the phase, toggling the availability of a single edge of $P_0$ or spoke only requires recomputing the DP tables along the path from its assigned cluster to the root, updating $O(\log n)$ tables in $O(1)$ time each. Since each graph update in $G_t$ toggles $O(1)$ edges and spokes in $J_t$, maintaining $T_t^J$ takes $O(\log n)$ time per update.

\begin{lemma}[Dynamic maintenance on $J_t$]\label{opt:dynamic}
The top-tree DP can be initialized at the start of a phase in $O(n\log n)$ time and $O(n)$ space, maintains the exact optimum $T_t^J$ minimizing~\eqref{eq:opt-triple} in $O(\log n)$ time per edge or spoke toggle, and supports:
\begin{enumerate}[nosep,label=(\roman*)]
  \item reading $\Lambda(T_t^J)$ in $O(1)$ time at the root;
  \item extracting the representative set $Q_t$ in $O(|Q_t|\log n)=O(B\log n)$ time; and
  \item reporting the edges of $F_t=T_t^J-h$ in $O(|E(F_t)|\log n)=O(n\log n)$ time.
\end{enumerate}
\end{lemma}
\begin{proof}
Constructing the top-tree hierarchy on $P_0$ and exposing $r_0$ takes $O(n\log n)$ time~\cite{AHLT05}, and initializing all DP tables takes $O(n)$ time and space by Lemma~\ref{opt:exact}. Each availability toggle affects only the $O(\log n)$ ancestors of the edge's or spoke's assigned cluster, each recomputed in $O(1)$ time, and $\Lambda(T_t^J)$ is stored at the root. 

To extract $Q_t$ without listing $F_t$, we inspect the optimal root entry: if its spoke count is zero, we return $Q_t=\{r_0\}$ in $O(1)$ time; otherwise, we follow its backpointers down the hierarchy, reporting locally selected spoke endpoints and recursing only into child entries with positive spoke count. Since each of the $|Q_t|\le\max\{1,|Z_t|\}=O(B)$ selected spokes is reached along a root-to-cluster path of length $O(\log n)$, this takes $O(|Q_t|\log n)=O(B\log n)$ time. 

Similarly, to report $E(F_t)$, we follow backpointers from the root and recurse only into child entries whose $P_0$-edge count (total edges minus spokes in~\eqref{eq:opt-triple}) is positive, taking $O(|E(F_t)|\log n)=O(n\log n)$ time.
\end{proof}

In our algorithm, we initialize the top-tree hierarchy at the beginning of each phase with all edges of $P_0$ enabled and all spokes disabled. Each graph update generates only $O(1)$ edge and spoke toggles, so maintaining the DP takes $O(\log n)$ time per update. Moreover, $J_t$ always contains a feasible Steiner tree: whenever an edge $uv$ of $P_0$ is disabled, the spokes $uh$ and $vh$ are enabled, providing the replacement path $u$--$h$--$v$. Replacing disabled edges in $P_0$ by these paths preserves connectivity among all terminals. Thus, a connected subgraph containing $R$ always exists in $J_t$.

\subsection{Reconnecting Surviving Components}
\label{sec:repair}
The forest $F_t=T_t^J-h$ already connects all original terminals within each of its connected components; by Lemma~\ref{opt:representatives}, reconnecting those components into a Steiner tree for $R$ reduces to connecting their representatives $Q_t$ in the current graph $G_t$. Running Mehlhorn's terminal-MST algorithm on $G_t$ or even on the current spanner $S_t$ after every update would be too slow for sublinear update time. Instead, we decouple per-update cost estimation from tree reporting:
\begin{enumerate}[nosep,label=(\roman*)]
  \item after each update, we run Mehlhorn's algorithm on the explicit HLS distance sparsifier $H_t$ (which has only $O(Bn^\eta)$ vertices and edges) to compute an MST $M_H$ of $K_{H_t}(Q_t)$ and update the reconnection cost estimate $\sigma\,w_{K_{H_t}(Q_t)}(M_H)$; and
  \item because $H_t$ is a non-subgraph sparsifier whose edges are virtual shortcuts rather than edges of $G_t$, whenever an explicit Steiner tree is requested we instead run Mehlhorn's construction and coherent expansion on the current dynamic spanner $S_t$ with terminal set $Q_t$ to obtain a tree $A_S\subseteq S_t$ connecting $Q_t$.
\end{enumerate}
Below, we first show how to maintain a designated vertex set in $H_t$ that always contains $Q_t$ while changing by only $O(1)$ vertices per graph update, and then describe how to estimate the reconnection cost on $H_t$ and materialize the reconnection tree on $S_t$.

\paragraph{Maintaining a larger designated set.}
Using $Q_t$ itself as the sparsifier's designated set would be costly: the representative set $Q_t$ extracted from $T_t^J$ may change substantially after each graph update, causing the top-tree DP to switch which components of $P_0-D_t$ connect to the hub, changing $\Omega(B)$ vertices of $Q_t$ at once. Instead, as the sparsifier's designated set, we use
\begin{equation}\label{eq:hls-designated}
U_t=Z_t\cup\{r_0\}
\end{equation}
containing the root terminal $r_0$ and all active vertices $Z_t$ accumulated so far in the current phase. Since every enabled spoke has its $P_0$-endpoint in $Z_t$ (and $Q_t=\{r_0\}$ when no spoke is used), we always have $Q_t\subseteq U_t$, so the distance guarantees of $H_t$ on $U_t$ automatically hold for all pairs in $Q_t$. Moreover, $U_t$ changes by only $O(1)$ vertices per graph update: we initialize $U_t=\{r_0\}$ at the start of the algorithm, insert the at most two newly active vertices into $U_t$ after each update in $O(n^\eta)$ amortized time (Theorem~\ref{thm:hls}), and remove the $O(B)$ active vertices from the ending phase at each phase boundary in $O(Bn^\eta)$ total time (or $O(n^\eta)$ amortized time per update). Throughout each phase, $|U_t|=O(B)$, so Theorem~\ref{thm:hls} guarantees
\[
|V(H_t)|+|E(H_t)|=O(|U_t|n^\eta)=O(Bn^\eta).
\]

\paragraph{Representative reconnection.}
After extracting $Q_t$ from the top-tree DP, we run Mehlhorn's construction (Theorem~\ref{thm:mehlhorn}) on the sparsifier $H_t$ with terminal set $Q_t$ (running multi-source Dijkstra from $Q_t$ rather than all of $U_t$ or $R$). This computes an exact MST $M_H$ of $K_{H_t}(Q_t)$ in $\Oh(|V(H_t)|+|E(H_t)|)=\Oh(Bn^\eta)$ deterministic time, and we store the cost upper bound $c_t=w(F_t)+\sigma\,w_{K_{H_t}(Q_t)}(M_H)$. When an explicit Steiner tree is requested, we run Mehlhorn's construction (Theorem~\ref{thm:mehlhorn}) on the current spanner $S_t$ with terminal set $Q_t$ to compute an MST $M_S$ of $K_{S_t}(Q_t)$, coherently expand $M_S$ along its owner forest in $S_t$ into a tree $A_S\subseteq S_t$ (Lemma~\ref{lem:coherent-expansion}), and in $F_t\cup A_S$ delete cycles and prune leaves outside $R$.

\begin{lemma}\label{lem:repair}
Maintaining the designated set $U_t=Z_t\cup\{r_0\}$ in the HLS sparsifier $H_t$ takes $O(n^\eta)$ amortized time per graph update and $O(Bn^\eta)$ amortized time for the $O(B)$ vertex deletions at each phase boundary, ensuring $Q_t\subseteq U_t$ and $|V(H_t)|+|E(H_t)|=O(Bn^\eta)$ throughout. After each update, computing the MST $M_H$ of $K_{H_t}(Q_t)$ and updating the cost estimate $c_t$ takes $\Oh(Bn^\eta)$ deterministic time.
\end{lemma}

\subsection{Approximation Analysis}
\label{sec:fd-approx}
We now turn to analyzing the approximation factor achieved by our algorithm against the optimum in the \emph{current} graph $G_t$. We bound the approximation factor in two steps. In Section~\ref{sec:fd-approx-forest}, we first show that the surviving forest $F_t$ has weight at most $2\sigma\OPT_{G_t}(R)$, and in Section~\ref{sec:fd-approx-repair} we show that the tree $A_S\subseteq S_t$ obtained by Mehlhorn's construction on the current spanner $S_t$ reconnects the component representatives $Q_t$ in $G_t$ at cost bounded by the scaled sparsifier MST weight $\sigma\,w_{K_{H_t}(Q_t)}(M_H)$ and proportional to $\OPT_{G_t}(R)$. Combining the surviving forest with this reconnection tree establishes our constant-factor approximation guarantee for integer weights (Theorem~\ref{thm:fd-phase-approx}), which Lemma~\ref{lem:rounding} then extends to polynomially bounded real weights.

\subsubsection{Bounding the Surviving Forest Cost}\label{sec:fd-approx-forest}
Consider a minimum Steiner tree $T^*$ connecting $R$ in $G_t$. Removing all newly inserted edges (i.e., those inserted during the current phase) from $T^*$ leaves components consisting only of surviving phase-start edges of $G^\circ$. Because hub spokes have weight zero in $J_t$ (their reconnection cost is paid in Section~\ref{sec:fd-approx-repair}), the weight of $T_t^J$ in $J_t$ equals the weight $w(F_t)$ of its $P_0$-edges. Moreover, since $T_t^J$ is a minimum-weight Steiner tree in $J_t$, we can upper-bound $w(F_t)$ by using the surviving components of $T^*$ to construct a feasible Steiner tree in $J_t$ of weight at most $2\sigma w(T^*)$.

The next lemma shows how to route each terminal-bearing component $K$ of $T^*$ in the frozen terminal metric $K_S(R)$, so that its terminals are connected both to one another and to an active terminal $a(x)\in A_t$, at total cost at most $2\sigma w(K)$. Rather than doubling all edges of $K$ in a closed tour, we traverse $K$ from an endpoint $x\in V(K)$ of a newly inserted edge to a terminal $r\in V(K)\cap R$ and use the weight along the $x$--$r$ path to pay for routing from $x$ to its nearest terminal $a(x)$ in $S$.

\begin{lemma}\label{lem:fd-component}
Suppose $T^*$ uses at least one newly inserted edge. Let $K$ be a terminal-containing component after all such edges are removed. Each such $K$ contains a vertex $x$ incident to a removed newly inserted edge, whose owner $a(x)$ in the frozen spanner $S$ has an enabled spoke $a(x)h$. The terminals $V(K)\cap R$, together with $a(x)$, are spanned by a connected subgraph in $K_S(R)$ of weight at most $2\sigma w(K)$.
\end{lemma}
\begin{proof}
First, such an endpoint $x$ must exist. Because $T^*$ is a tree and at least one newly inserted edge is removed, at least two components remain. Each component $K$ was connected to the other components only through removed newly inserted edges, so it contains at least one vertex $x$ incident to such an edge. The owner $a(x)$ of $x$ in the frozen spanner $S$ was added to $A_t$ when that edge was inserted, so the spoke $a(x)h$ is available.

Next, we fix any terminal $r\in V(K)\cap R$. Since $K$ is a component obtained by deleting edges from the tree $T^*$, it is also a tree; let $d_K(x,r)$ be the length of its unique $x$--$r$ path. Starting at $x$, we progress along this path: at each path vertex, we first explore every subtree attached outside the path by a depth-first traversal, returning to that vertex before advancing to the next path vertex, and at $r$ we finish exploring its attached subtrees and stop. This walk, denoted by $W$, visits every vertex of $K$, traversing each path edge once and every other edge twice, as illustrated in Figure~\ref{fig:fd-open-walk}. Its length is
\[
w(W)=2w(K)-d_K(x,r).
\]

\begin{figure}[tb]
\centering
\begin{tikzpicture}[
  every node/.style={font=\small},
  vertex/.style={circle,fill=black,inner sep=1.6pt},
  tree edge/.style={black!80,line width=0.8pt},
  owner route/.style={dash pattern=on 3pt off 2.7pt,line width=0.85pt},
  spoke/.style={dash pattern=on 3pt off 2.7pt,line width=0.85pt},
  walk/.style={red!72!black,line width=0.9pt,-{Stealth[length=1.7mm]}},
  other/.style={blue!65!black},
  line cap=round,line join=round,
  >=Stealth
]
\coordinate (x) at (-0.75,-0.25);
\coordinate (p) at (-2.05,-0.9);
\coordinate (r) at (-3.2,-1.85);
\coordinate (b5) at (-4.45,-2.0);
\coordinate (b1) at (-2.6,0.4);
\coordinate (b2) at (-0.85,-1.5);
\coordinate (b4) at (-3.2,-3.0);
\draw[blue!65!black,line width=1.15pt] (x) -- (p) -- (r);
\foreach \u/\v in {p/b1,p/b2,r/b4,r/b5}
  \draw[tree edge] (\u) -- (\v);
\node[blue!65!black,font=\footnotesize,align=left,anchor=west] at (0.05,-1.2) {$x$--$r$ path:\\length $d_K(x,r)$};
\draw[blue!65!black,thin] (-0.08,-1.05) to[out=155,in=-25] (-1.45,-0.6);

\draw[walk] (x) to[bend left=11] (p);
\draw[walk] (p) to[bend left=11] (r);
\foreach \u/\v in {p/b1,p/b2,r/b4,r/b5} {
  \draw[walk] (\u) to[bend left=17] (\v);
  \draw[walk] (\v) to[bend left=17] (\u);
}
\foreach \v in {x,p,r,b5,b1,b2,b4}
  \node[vertex] at (\v) {};
\node[below right=1pt] at (x) {$x$};
\node[above left=2pt] at (r) {$r$};
\node at (-4.5,-1.3) {$K$};
\node[red!72!black,align=center,font=\footnotesize] at (-4.0,1.65) {Explore a subtree,\\then return};
\draw[red!72!black,thin,->] (-3.25,1.32) to[out=-35,in=125] (-2.75,0.45);
\node[red!72!black,align=center,font=\footnotesize] at (-0.6,-2.55) {Then advance\\toward $r$};
\draw[red!72!black,thin,->] (-1.35,-2.2) to[out=150,in=-20] (-2.55,-1.35);

\node[vertex,label=above:$h$] (h) at (0.9,2.2) {};
\node[vertex,label=above left:$a(x)$] (ax) at (-0.4,1.15) {};
\node[vertex,other,label={[other]above right:$a(y)$}] (ay) at (2.2,1.15) {};
\node[vertex,other,label=below left:$y$] (y) at (2.55,-0.25) {};
\draw[owner route] (x) .. controls (-1.2,0.35) and (-0.85,0.85) .. node[left=2pt,font=\footnotesize,align=center,pos=0.82] {shortest path\\in $S$} (ax);
\draw[spoke] (ax) -- node[above left=1pt,pos=0.55] {$0$} (h);
\draw[spoke,other] (h) -- node[above right=1pt,pos=0.45] {$0$} (ay);
\draw[owner route,other] (ay) .. controls (2.65,0.85) and (3.0,0.35) .. node[right=2pt,font=\footnotesize,align=center,pos=0.22] {shortest path\\in $S$} (y);

\draw[other,line width=0.8pt,dashed] (x) -- node[above=3pt,font=\footnotesize,align=center] {newly inserted edge $xy$\\(removed)} (y);
\draw[other,line width=0.9pt] (3.55,-0.25) ellipse (1.0 and 0.65);
\node[other] at (3.55,-0.2) {$K'$};
\node[vertex,other,label={[other]right:$r'$}] at (3.9,-0.5) {};
\node[other,font=\footnotesize,align=left,anchor=north west] at (4.25,-0.85) {another\\component};
\end{tikzpicture}
\caption{After removing the newly inserted edge $xy$, its endpoints lie in surviving components $K$ and $K'$, and their frozen owners have enabled zero-weight spokes to $h$. In $K$, the unique $x$--$r$ path is highlighted in blue and has length $d_K(x,r)$. At each vertex on the blue path, the red walk explores the off-path subtrees and returns to that vertex before moving to the next vertex. At $r$, it explores the remaining subtrees, returns to $r$, and stops. Path edges are traversed once and all other edges twice, giving length $2w(K)-d_K(x,r)$.}
\label{fig:fd-open-walk}
\end{figure}
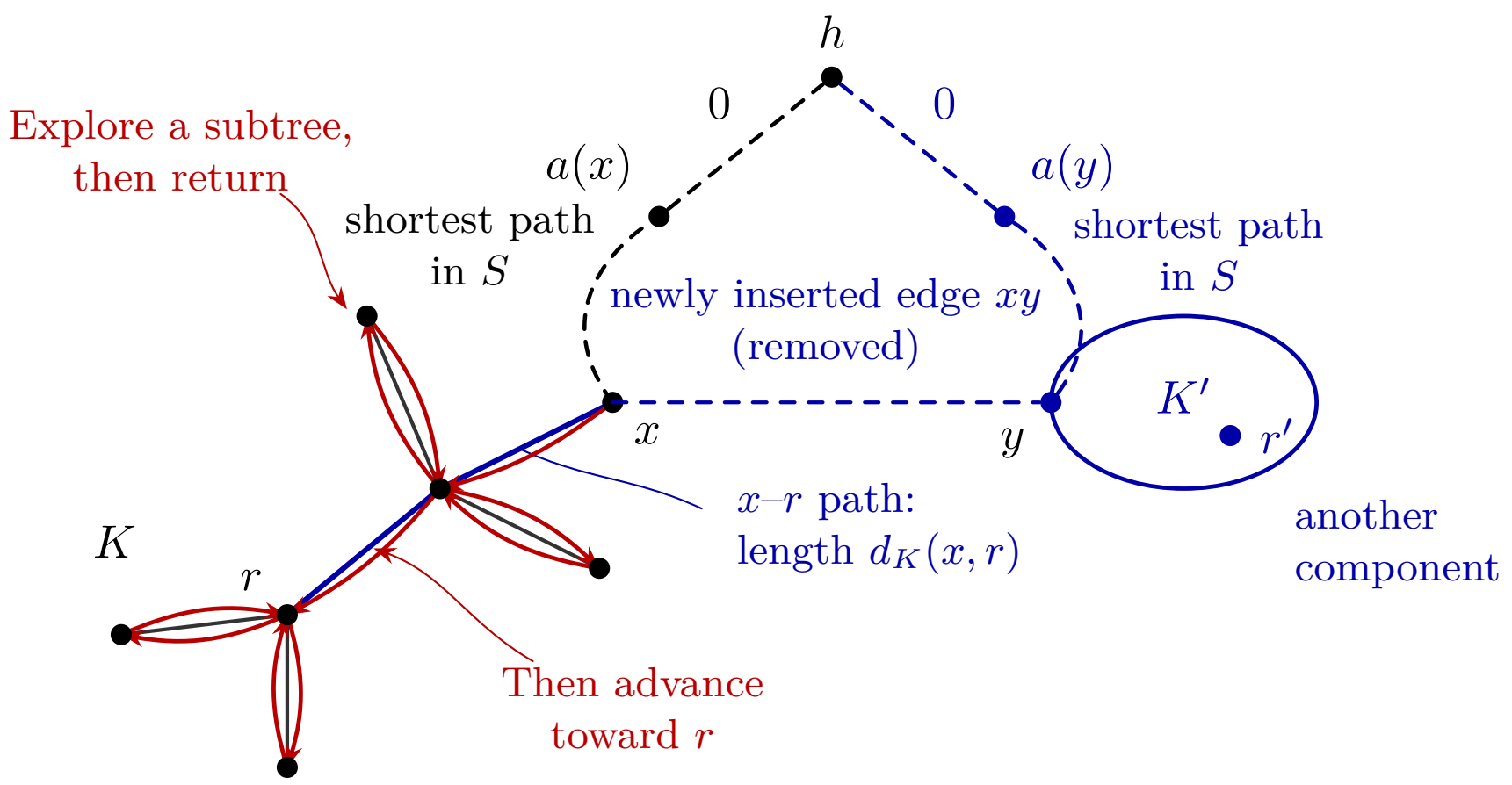

Every edge of $W$ survives from $G^\circ$. For each traversal of an edge $uv$ in $W$, we replace it by a shortest $u$--$v$ path in the frozen spanner $S$, which has length
\begin{align*}
d_S(u,v)
&\le \sigma d_{G^\circ}(u,v)
&&\text{($S$ is a $\sigma$-spanner of $G^\circ$)}\\
&\le \sigma w(uv)
&&\text{($uv$ itself is a path in $G^\circ$)}.
\end{align*}
Let $W_S$ be the resulting walk in $S$, obtained by concatenating the shortest paths that replace successive edge traversals of $W$. Summing over the edge traversals of $W$, with repetitions, gives
\begin{align*}
w(W_S)
&=\sum_{uv\in W}d_S(u,v) \le \sigma\sum_{uv\in W}w(uv) =\sigma\bigl(2w(K)-d_K(x,r)\bigr).
\end{align*}

The shortest path from $a(x)$ to $x$ in the frozen spanner $S$ has length
\begin{align*}
\delta(x)
&=d_S(a(x),x)\\
&\le d_S(r,x)
&&\text{($a(x)$ is the nearest terminal to $x$; $r\in R$)}\\
&\le \sigma d_{G^\circ}(r,x)
&&\text{($S$ is a $\sigma$-spanner of $G^\circ$)}\\
&\le \sigma d_K(r,x)
&&\text{($K\subseteq G^\circ$)}.
\end{align*}
Concatenating this shortest $a(x)$--$x$ path in $S$ with $W_S$ yields a walk in $S$ that begins at $a(x)\in R$ and visits every terminal in $V(K)\cap R$, of total length at most
\begin{equation}\label{eq:fd-open-tour}
\delta(x)+w(W_S)\le\delta(x)+\sigma\bigl(2w(K)-d_K(x,r)\bigr)\le 2\sigma w(K).
\end{equation}
The subtraction in the $x$--$r$ walk pays for the shortest path from $a(x)$ to $x$ in $S$.

Finally, since this walk in $S$ starts at the terminal $a(x)\in R$ and visits all terminals in $V(K)\cap R$, we list the visited terminals of $(V(K)\cap R)\cup\{a(x)\}$ in order of visit (ignoring consecutive repetitions of the same terminal) and replace the walk segment between each consecutive pair $u,v$ by the metric edge $uv$ of weight $d_S(u,v)$ in $K_S(R)$. Because $d_S(u,v)$ is at most the length of the walk segment in $S$ between $u$ and $v$, the resulting subgraph in $K_S(R)$ spans $(V(K)\cap R)\cup\{a(x)\}$ and has weight at most~\eqref{eq:fd-open-tour}.
\end{proof}

Lemma~\ref{lem:fd-component} connects the terminals of each surviving component $K$ of $T^*$ to an active terminal $a(x_K)$ using edges of the full terminal metric $K_S(R)$, and the enabled spoke $a(x_K)h$ links all these active terminals through the hub $h$. To turn this into a valid Steiner tree inside $J_t$, we use the MST augmentation property (Lemma~\ref{lem:mst-augmentation}) to replace the clique edges of $K_S(R)$ with edges of the phase-start terminal MST $T_0$, coherently expand those $T_0$ edges into paths in $P_0$ (Lemma~\ref{lem:coherent-expansion}), and bypass any deleted edge $uv\in D_t$ of $P_0$ via the zero-weight spokes $u$--$h$--$v$.

\begin{lemma}\label{lem:fd-forest-cost}
At any step $t$ of a phase, the surviving forest $F_t=T_t^J-h$ satisfies
\[
w(F_t)\le 2\sigma\OPT_{G_t}(R).
\]
\end{lemma}
\begin{proof}
First suppose that $T^*$ contains at least one newly inserted edge. For each terminal-containing component $K$ of $T^*$ after removing newly inserted edges, Lemma~\ref{lem:fd-component} provides a connected subgraph in $K_S(R)$ spanning $(V(K)\cap R)\cup\{a(x_K)\}$ of weight at most $2\sigma w(K)$ together with an enabled spoke $a(x_K)h$. Taking the union of these subgraphs and the set $E_h$ of corresponding spokes $a(x_K)h$ over all such components $K$ yields a connected subgraph $L\subseteq K_S(R)\cup E_h$ spanning $R\cup\{h\}$. Since spokes have weight zero in $J_t$ and the components $K$ are edge-disjoint subgraphs of $T^*$, we have
\[
w(L)\le \sum_K 2\sigma w(K) =2\sigma\sum_K w(K)\le 2\sigma w(T^*).
\]
Because $T_0$ is an MST of $K_S(R)$ and $L$ is a connected subgraph of $K_S(R)\cup E_h$ spanning $R\cup\{h\}$, Lemma~\ref{lem:mst-augmentation} implies that an MST $M$ of $T_0\cup E_h$ satisfies
\[
w(M) = \mst(T_0\cup E_h) = \mst(K_S(R)\cup E_h) \le w(L) \le 2\sigma w(T^*).
\]
Now we map $M$ into the auxiliary graph $J_t$. By Lemma~\ref{lem:coherent-expansion}, replacing the $T_0$-edges of $M$ by their coherent routes in $P_0$ (while keeping the spokes of $M$ in $E_h$) produces a connected subgraph of $P_0\cup E_h$ spanning $R$ of weight at most $w(M)$. If any edge $uv\in E(P_0)$ used by this subgraph was deleted during the phase ($uv\in D_t$), then both endpoints $u,v\in\operatorname{end}(D_t)\subseteq Z_t$ have enabled zero-weight spokes $uh$ and $vh$ in $J_t$; replacing each such deleted edge $uv\in D_t$ by the two-spoke path $u$--$h$--$v$ preserves connectivity on $R$ without increasing weight. Removing cycles from the resulting connected subgraph yields a feasible Steiner tree $Y$ in $J_t$ with $w(Y)\le w(M)\le 2\sigma w(T^*)$.

In the remaining case where $T^*$ contains no newly inserted edge, we have $T^*\subseteq G^\circ$. By the metric-MST bound (Lemma~\ref{lem:metric-mst}) and the fact that $S$ is a $\sigma$-spanner of $G^\circ$,
\[
w(T_0)\le 2\OPT_S(R)\le 2\sigma\OPT_{G^\circ}(R)\le 2\sigma w(T^*).
\]
Expanding $T_0$ into $P_0$ and replacing any deleted edge $uv\in D_t$ by the zero-weight path $u$--$h$--$v$ as above again yields a feasible Steiner tree $Y$ in $J_t$ with $w(Y)\le w(T_0)\le 2\sigma w(T^*)$.

In either case, since $T_t^J$ minimizes weight among all Steiner trees connecting $R$ in $J_t$ and its spokes have weight zero, removing the spokes from $T_t^J$ leaves $F_t$ with
\[
w(F_t)=w(T_t^J)\le w(Y)\le 2\sigma w(T^*)=2\sigma\OPT_{G_t}(R).\qedhere
\]
\end{proof}

\subsubsection{Bounding the Representative Reconnection Cost}\label{sec:fd-approx-repair}
We now bound the weight of the tree $A_S\subseteq S_t$ (and the stored estimate $\sigma\,w_{K_{H_t}(Q_t)}(M_H)$) used to reconnect the surviving components of $F_t$ through their representatives $Q_t$. Because $Q_t$ may include non-terminal vertices of $P_0$, we first bound the optimal cost $\OPT_{G_t}(Q_t)$ of connecting $Q_t$ in $G_t$ in terms of $\OPT_{G_t}(R)$ (Lemma~\ref{lem:fd-representative-opt}), and then show that $A_S$ and $\sigma\,w_{K_{H_t}(Q_t)}(M_H)$ connect $Q_t$ within a $2\sigma\rho_\eta$ factor.

\begin{lemma}\label{lem:fd-representative-opt}
At any step $t$ of a phase, the minimum Steiner tree weight for the representative set $Q_t$ in $G_t$ satisfies
\[
\OPT_{G_t}(Q_t)\le w(F_t)+\OPT_{G_t}(R)
\le(1+2\sigma)\OPT_{G_t}(R).
\]
\end{lemma}
\begin{proof}
By Lemma~\ref{opt:representatives}, each representative in $Q_t$ belongs to a component of $F_t$ that also contains a terminal in $R$. Consequently, the union of $F_t$ with a minimum Steiner tree for $R$ in $G_t$ connects all vertices of $Q_t$ and has weight at most $w(F_t)+\OPT_{G_t}(R)$, proving the first inequality; Lemma~\ref{lem:fd-forest-cost} then gives the second.
\end{proof}

Having related $\OPT_{G_t}(Q_t)$ to $\OPT_{G_t}(R)$, we next bound both the scaled sparsifier MST weight $\sigma\,w_{K_{H_t}(Q_t)}(M_H)$ and the weight of the tree $A_S\subseteq S_t$ obtained by coherently expanding the terminal-metric MST $M_S$ of $K_{S_t}(Q_t)$.

\begin{lemma}\label{lem:fd-repair-cost}
At any step $t$ of a phase, the tree $A_S\subseteq S_t$ connects all representatives in $Q_t$ in $G_t$ and satisfies
\[
w(A_S)\le \sigma\,w_{K_{H_t}(Q_t)}(M_H)\le 2\sigma\rho_\eta\OPT_{G_t}(Q_t).
\]
\end{lemma}
\begin{proof}
    Let $M_G$ be an MST of $K_{G_t}(Q_t)$, used only for analysis. Since both $K_{H_t}(Q_t)$ and $K_{G_t}(Q_t)$ are complete graphs on $Q_t$, the edge set of $M_G$ also forms a spanning tree of $K_{H_t}(Q_t)$, where each edge $uv\in E(M_G)$ has weight $d_{H_t}(u,v)$ instead of $d_{G_t}(u,v)$. Therefore,
    \begin{align}
    w_{K_{H_t}(Q_t)}(M_H)
    &\le w_{K_{H_t}(Q_t)}(M_G)
    &&\text{($M_H$ is an MST of $K_{H_t}(Q_t)$)}\notag\\
    &=\sum_{uv\in E(M_G)}d_{H_t}(u,v)\notag\\
    &\le\rho_\eta\sum_{uv\in E(M_G)}d_{G_t}(u,v)
    &&\text{(sparsifier stretch, Theorem~\ref{thm:hls}, since $Q_t\subseteq U_t$)}\\
    &=\rho_\eta w_{K_{G_t}(Q_t)}(M_G)\notag\\
    &\le2\rho_\eta\OPT_{G_t}(Q_t)
    &&\text{(the metric-MST bound, Lemma~\ref{lem:metric-mst})}.
    \label{eq:fd-metric-repair}
    \end{align}
    Similarly, since $M_S$ is an MST of $K_{S_t}(Q_t)$ and $M_H$ is a spanning tree on the same vertex set $Q_t$, expanding $M_S$ coherently in $S_t$ (Lemma~\ref{lem:coherent-expansion}) produces a tree $A_S\subseteq S_t\subseteq G_t$ connecting $Q_t$ with
    \begin{align}
    w(A_S)
    &\le w_{K_{S_t}(Q_t)}(M_S)
    &&\text{(the union counts shared edges only once, Lemma~\ref{lem:coherent-expansion})}\notag\\
    &\le w_{K_{S_t}(Q_t)}(M_H)
    &&\text{($M_S$ is an MST of $K_{S_t}(Q_t)$)}\notag\\
    &=\sum_{uv\in E(M_H)}d_{S_t}(u,v)\notag\\
    &\le\sigma\sum_{uv\in E(M_H)}d_{G_t}(u,v)
    &&\text{(spanner stretch, Theorem~\ref{thm:cp-spanner})}\\
    &\le\sigma\sum_{uv\in E(M_H)}d_{H_t}(u,v)
    &&\text{(noncontraction, Theorem~\ref{thm:hls})}\notag\\
    &=\sigma\,w_{K_{H_t}(Q_t)}(M_H)\notag\\
    &\le2\sigma\rho_\eta\OPT_{G_t}(Q_t)
    &&\text{(Inequality~\eqref{eq:fd-metric-repair})}.\qedhere
    \end{align}
\end{proof}

Combining the bound on the surviving forest $F_t$ (Lemma~\ref{lem:fd-forest-cost}) with the bounds on the reconnection tree $A_S$ and stored cost estimate $c_t$ (Lemmas~\ref{lem:fd-representative-opt} and~\ref{lem:fd-repair-cost}) yields our approximation guarantee for integer-weighted graph $G_t$.

\begin{theorem}\label{thm:fd-phase-approx}
For positive polynomially bounded integer edge weights, the Steiner tree $T_t$ maintained by the algorithm in $G_t$ and the stored cost estimate $c_t=w(F_t)+\sigma\,w_{K_{H_t}(Q_t)}(M_H)$ satisfy
\[
\OPT_{G_t}(R)\le w(T_t)\le c_t\le\bigl[2\sigma+2\sigma\rho_\eta(1+2\sigma)\bigr]\OPT_{G_t}(R).
\]
\end{theorem}
\begin{proof}
The union $F_t\cup A_S$ connects all original terminals. Removing cycles to obtain a spanning tree and then pruning leaves outside $R$ gives a Steiner tree $T_t$ for $R$ satisfying
\begin{align*}
\OPT_{G_t}(R)\le w(T_t)
&\le w(F_t)+w(A_S)\\
&\le w(F_t)+\sigma\,w_{K_{H_t}(Q_t)}(M_H) = c_t
&&\text{(Lemma~\ref{lem:fd-repair-cost})}\\
&\le2\sigma\OPT_{G_t}(R)+2\sigma\rho_\eta\OPT_{G_t}(Q_t)
&&\text{(Lemmas~\ref{lem:fd-forest-cost} and~\ref{lem:fd-repair-cost})}\\
&\le2\sigma\OPT_{G_t}(R)+2\sigma\rho_\eta(1+2\sigma)\OPT_{G_t}(R)
&&\text{(Lemma~\ref{lem:fd-representative-opt})}\\
&=\bigl[2\sigma+2\sigma\rho_\eta(1+2\sigma)\bigr]\OPT_{G_t}(R).&&\qedhere
\end{align*}
\end{proof}

Thus, $c_t$ upper-bounds the reported tree's cost and is a constant-factor approximation to the current optimum.

\paragraph{Returning to real weights.}
\label{sec:rounding}
We now transfer the approximation guarantee from the rounded integer weights $\widehat w(e)=\lceil w(e)/\varepsilon\rceil$ (for fixed $\varepsilon\in(0,1)$) to the original real weights $w(e)\in[1,W]$ at the cost of a $(1+\varepsilon)$ factor.

Since $\lceil x\rceil\in[x,x+1)$ and $w(e)\ge 1$, every edge $e$ satisfies
\begin{equation}\label{eq:rounding-edge-bound}
w(e) \le \varepsilon\left\lceil\frac{w(e)}{\varepsilon}\right\rceil = \varepsilon\widehat w(e) \le \varepsilon\left(\frac{w(e)}{\varepsilon}+1\right) = w(e)+\varepsilon \le (1+\varepsilon)w(e).
\end{equation}
For fixed $\varepsilon$, the maximum integer weight $\lceil W/\varepsilon\rceil$ remains polynomial in $n$. Moreover, because rounding only rescales edge weights, both instances share the exact same set of feasible Steiner trees.

\begin{lemma}\label{lem:rounding}
A $C$-approximation in the integer-weighted instance gives a $C(1+\varepsilon)$-approximation in the original real-weighted instance. Rounding takes $O(m_0)$ initial time and $O(1)$ additional time per update.
\end{lemma}
\begin{proof}
Summing~\eqref{eq:rounding-edge-bound} over the edges of any subgraph $H$ gives $w(H)\le \varepsilon\widehat w(H)\le (1+\varepsilon)w(H)$. Let $T$ be the $C$-approximate Steiner tree returned in the integer instance (with optimum $\widehat{\OPT}$), and let $T^*$ be an optimal Steiner tree under the original weights $w$. Since $T^*$ is also a feasible Steiner tree in the integer instance, $\widehat{\OPT}\le\widehat w(T^*)$, and therefore
\[
  w(T) \le \varepsilon\widehat w(T) \le \varepsilon C\widehat{\OPT} \le \varepsilon C\widehat w(T^*) \le C(1+\varepsilon)w(T^*) = C(1+\varepsilon)\OPT_{G_t}(R).
\]
Evaluating $\widehat w(e)=\lceil w(e)/\varepsilon\rceil$ takes $O(1)$ time per edge, giving $O(m_0)$ initial time and $O(1)$ time per edge update.
\end{proof}

Taking $C=2\sigma+2\sigma\rho_\eta(1+2\sigma)$, we get approximation factor $(1+\varepsilon)\bigl[2\sigma+2\sigma\rho_\eta(1+2\sigma)\bigr]$.

Under the original real weights, return $\varepsilon c_t$ for a cost-estimation query; this upper-bounds the reported tree's cost and satisfies the same $C(1+\varepsilon)$ approximation guarantee.

\subsection{Time and Space Analysis}
\label{sec:fd-runtime}
We now analyze the running time and space usage of our fully dynamic algorithm (Algorithm~\ref{alg:fd-initialize}) for fixed approximation parameters $\nu,\eta,\varepsilon\in(0,1)$ with $\nu+\eta<1$. In Section~\ref{sec:fd-reporting}, we first bound the running time of each of the four procedures (initialization, phase reconstruction, single-update processing, and tree reporting) in terms of the phase length $B$, and then choose $B$ to balance their running times. We bound the space usage in Lemma~\ref{lem:fd-space}.

\subsubsection{Running Time Analysis}
\label{sec:fd-reporting}

\begin{lemma}\label{lem:fd-routine-times}
For fixed parameters $\nu,\eta,\varepsilon\in(0,1)$ and phase length $B\ge 1$, the procedures of Algorithm~\ref{alg:fd-initialize} satisfy the following deterministic time bounds:
\begin{enumerate}[nosep,label=(\roman*)]
  \item \emph{Initialization:} $O\bigl((m_0+n)n^{\max\{\nu,\eta\}}\bigr)$ time.
  \item \emph{Phase reconstruction:}  $\Oh(n^{1+\nu}/B)+O(n^\eta)$ amortized update time over $B$ updates.
  \item \emph{Single-update processing:} $O(n^\nu)+\Oh(Bn^\eta)$ amortized time per update.
  \item \emph{Tree reporting:} $\Oh(n^{1+\nu})$ worst-case time per query.
\end{enumerate}
\end{lemma}
\begin{proof}
We establish the four bounds in order.

\paragraph{Initialization.}
Rounding the $m_0$ initial edge weights takes $O(m_0)$ time (Lemma~\ref{lem:rounding}), initializing the dynamic spanner $S_0$ takes $O((m_0+n)n^\nu)$ time (Theorem~\ref{thm:cp-spanner}), and initializing the HLS sparsifier $H_0$ with designated set $U_0=\{r_0\}$ takes $O((m_0+n)n^\eta)$ time (Theorem~\ref{thm:hls}). For the initial phase reconstruction on $G_0$, the frozen spanner snapshot $S\subseteq G_0$ has at most $m_0$ edges and no active vertices need to be removed from a prior phase, so $\textsc{StartPhase}$ takes $\Oh(m_0+n)$ time as shown next, which is absorbed by $O\bigl((m_0+n)n^{\max\{\nu,\eta\}}\bigr)$.

\paragraph{Phase reconstruction.}
Let $M=|E(S)|$ be the number of edges in the frozen spanner snapshot $S$ at the start of a phase, where $M=O(n^{1+\nu})$ by Theorem~\ref{thm:cp-spanner}. Copying $S$ and running Mehlhorn's construction to compute the owner forest and terminal-metric MST $T_0$ takes $\Oh(M+n)=\Oh(n^{1+\nu})$ time (Theorem~\ref{thm:mehlhorn}), coherently expanding $T_0$ into the tree $P_0$ takes $O(n)$ time (Lemma~\ref{lem:coherent-expansion}), and building the top-tree hierarchy on $P_0$ and initializing its DP tables takes $O(n\log n)$ time (Lemmas~\ref{opt:exact} and~\ref{opt:dynamic}). Resetting the sparsifier's designated set $U_t$ to $\{r_0\}$ removes the $O(B)$ active vertices from the preceding phase in $O(Bn^\eta)$ amortized time (Lemma~\ref{lem:repair}). Charging each rebuild after the first to the $B$ updates of the preceding phase gives $\Oh(n^{1+\nu}/B)+O(n^\eta)$ amortized time per update.

\paragraph{Single-update processing.}
On each edge update, rounding its weight takes $O(1)$ time (Lemma~\ref{lem:rounding}), updating the dynamic spanner $S_t$ takes $O(n^\nu)$ amortized time (Theorem~\ref{thm:cp-spanner}), updating the top-tree DP on $J_t$ for the $O(1)$ toggled edges and spokes takes $O(\log n)$ time, and extracting the representative set $Q_t$ takes $O(B\log n)$ time (Lemma~\ref{opt:dynamic}). Forwarding the edge update to $H_t$ and inserting the $O(1)$ newly active vertices into $U_t$ takes $O(n^\eta)$ amortized time (Theorem~\ref{thm:hls} and Lemma~\ref{lem:repair}), and running Mehlhorn's construction on $(H_t,Q_t)$ to update the cost estimate $c_t$ takes $\Oh(Bn^\eta)$ time (Lemma~\ref{lem:repair}). Summing these costs gives $O(n^\nu)+\Oh(Bn^\eta)$ amortized time per update.

\paragraph{Tree reporting.}
When an explicit Steiner tree is requested, recovering the edges of the forest $F_t$ via the top-tree DP backpointers takes $O(|E(F_t)|\log n)=O(n\log n)$ time (Lemma~\ref{opt:dynamic}). Running Mehlhorn's construction on the current spanner $S_t$ with terminal set $Q_t$ and coherently expanding the resulting terminal-metric MST $M_S$ into the tree $A_S\subseteq S_t$ takes $\Oh(|E(S_t)|+n)=\Oh(n^{1+\nu})$ worst-case time (Theorems~\ref{thm:cp-spanner} and~\ref{thm:mehlhorn}, and Lemma~\ref{lem:coherent-expansion}). Since $F_t$ and $A_S$ have $O(n)$ edges in total, removing cycles and pruning non-terminal leaves in $F_t\cup A_S$ takes $O(n)$ time, giving $\Oh(n^{1+\nu})$ worst-case reporting time.
\end{proof}
Thus, any sequence of $U\ge 1$ updates incurs total update time $\Oh(U(n^\nu+Bn^\eta) + \lfloor U/B\rfloor n^{1+\nu})$ beyond initialization, corresponding to an amortized update time of
\begin{equation}\label{eq:fd-total-B}
T_{\mathrm{upd}}(B) = \Oh\!\left(\frac{n^{1+\nu}}{B}+n^\nu+Bn^\eta\right).
\end{equation}
We balance the first and third terms by setting
\begin{equation}\label{eq:fd-B}
B=\left\lceil n^{(1+\nu-\eta)/2}\right\rceil,
\end{equation}
which (since $\nu<1$ and $\eta>0$ imply $(1+\nu+\eta)/2>\nu$) gives deterministic amortized update time $T_{\mathrm{upd}}=\Oh\!\left(n^{(1+\nu+\eta)/2}\right)$.

\begin{lemma}\label{lem:fd-space}
Let $m_t=|E(G_t)|$ be the current edge count. For fixed constants $\nu,\eta\in(0,1)$ with $\nu+\eta<1$ and our choice of $B$ in~\eqref{eq:fd-B}, the total space used is
\[
  O\bigl(m_t n^{\max\{\nu,\eta\}} + n^{1+\max\{\nu,\eta\}}\bigr).
\]
\end{lemma}
\begin{proof}
Storing the current graph takes $O(m_t+n)$ space; maintaining the dynamic spanner $S_t$, its frozen snapshot $S$, and Mehlhorn's workspace on $S$ and $S_t$ takes $O(m_t n^\nu+n^{1+\nu})$ space (Theorems~\ref{thm:cp-spanner} and~\ref{thm:mehlhorn}); maintaining the HLS sparsifier $H_t$ takes $O((m_t+n)n^\eta)$ space (Theorem~\ref{thm:hls}); and storing the owner forest, $T_0$, $P_0$, and the top-tree hierarchy with its DP tables takes $O(n)$ space (Theorem~\ref{thm:mehlhorn} and Lemma~\ref{opt:dynamic}). Within a phase, maintaining $A_t,D_t,Z_t,U_t,Q_t$ and Mehlhorn's workspace on $H_t$ takes $O(Bn^\eta)$ space (Theorem~\ref{thm:mehlhorn} and Lemma~\ref{lem:repair}). Summing these bounds gives total space
\begin{equation*}
  O\bigl(m_t n^{\max\{\nu,\eta\}} + n^{1+\max\{\nu,\eta\}} + Bn^\eta\bigr).
\end{equation*}
For $B=\lceil n^{(1+\nu-\eta)/2}\rceil$ and $\nu+\eta<1$, we have $Bn^\eta=O(n^{(1+\nu+\eta)/2})=o(n)$, leaving $O\bigl(m_t n^{\max\{\nu,\eta\}} + n^{1+\max\{\nu,\eta\}}\bigr)$.\qedhere
\end{proof}

Combining Theorem~\ref{thm:fd-phase-approx} and Lemma~\ref{lem:rounding} with Lemmas~\ref{lem:fd-routine-times} and~\ref{lem:fd-space} yields the full parameterized guarantee for our fully dynamic algorithm.

\begin{theorem}[Parameterized version of Theorem~\ref{thm:main-fully-dynamic}]\label{thm:fd-full}
Let $G$ be a connected $n$-vertex fully dynamic undirected graph with $m_0$ edges initially, a terminal set $R$, and polynomially bounded positive edge weights over all updates. For any fixed constants $\nu,\eta,\varepsilon\in(0,1)$ with $\nu+\eta<1$, setting $\sigma=\sigma_\nu$ and $B=\bigl\lceil n^{(1+\nu-\eta)/2}\bigr\rceil$ gives a deterministic fully dynamic algorithm (against an adaptive adversary) that maintains a $(1+\varepsilon)\bigl[2\sigma+2\sigma\rho_\eta(1+2\sigma)\bigr]$-approximate Steiner tree for $R$ in $G$ (where $\sigma_\nu=2^{\operatorname{poly}(1/\nu)}$ and $\rho_\eta=2^{\operatorname{poly}(1/\eta)}$ are the stretch factors of Theorems~\ref{thm:cp-spanner} and~\ref{thm:hls}) with:
\begin{itemize}[nosep]
  \item $\Oh\!\left(n^{(1+\nu+\eta)/2}\right)$ amortized update time,
  \item $O\bigl((m_0+n)n^{\max\{\nu,\eta\}}\bigr)$ initialization time, and
  \item $O\bigl(m_t n^{\max\{\nu,\eta\}}+n^{1+\max\{\nu,\eta\}}\bigr)$ space, where $m_t$ is the current number of edges.
\end{itemize}
After every update, cost queries take $O(1)$ worst-case time, and reporting an explicit Steiner tree takes $\Oh(n^{1+\nu})$ worst-case time.
\end{theorem}
\begin{proof}
Approximation for both the reported tree and the $O(1)$-time cost estimate $\varepsilon c_t$ follows from Theorem~\ref{thm:fd-phase-approx} and Lemma~\ref{lem:rounding}. Initialization, amortized update, and worst-case reporting times follow from Lemma~\ref{lem:fd-routine-times} and~\eqref{eq:fd-total-B}--\eqref{eq:fd-B}, and space follows from Lemma~\ref{lem:fd-space}.

In particular, for any fixed $0<\zeta<1/2$, choosing $\nu=\eta=\zeta$ and $\varepsilon=1/2$ yields $\Oh(n^{1/2+\zeta})$ amortized update time, $O((m_0+n)n^\zeta)$ initialization time, $O(m_t n^\zeta+n^{1+\zeta})$ space, and $\Oh(n^{1+\zeta})$ worst-case reporting time, proving Theorem~\ref{thm:main-fully-dynamic}.
\end{proof}

\begin{remark}[Edge-weight changes]\label{rem:weight-changes}
To update the weight of an edge $uv$, we simply delete $uv$ and immediately reinsert it with its new weight before running the top-tree extraction and Mehlhorn reconnection in \textsc{Update}. Deleting $uv$ first avoids creating parallel edges, and even if $uv$ is a bridge so that the graph is momentarily disconnected between the two internal operations, the underlying spanner and HLS sparsifier support disconnected graphs without issue, while Mehlhorn's reconnection on $(H_t,Q_t)$ only runs once $uv$ is back in place. In $J_t$, the deletion disables $uv$ if it belonged to $P_0$ (so $F_t$ never uses a stale phase-start weight) and enables its endpoint spokes $uh,vh$, while the insertion enables the owner spokes $a(u)h,a(v)h$ to account for the new weight, increasing the per-update work and $|Z_t|,|U_t|=O(B)$ by at most a constant factor.
\end{remark}

\appendix
\section{Insertion-only Implementation and Analysis}
\label{app:insertion-only}

This appendix provides the implementation details and analysis of the insertion-only algorithm in Section~\ref{sec:insertion-only}. We begin in Appendix~\ref{app:insertion-only-impl} with the maintenance of our Steiner augmentation structures: the distance and next-hop matrices of the auxiliary multigraph $L_F$ on the endpoint set $X$, and the augmented terminal MST $\mathcal M$ on $T_0\cup C_F$. We then establish the approximation guarantee in Appendix~\ref{inc:approximation} and analyze the running time and space usage in Appendix~\ref{inc:runtime}. For fixed integers $s,r\ge 2$ and constant $0<\delta<1$, $\sigma=(2s-1)(1+\delta)$ and $A=2r-1$ denote the spanner stretch (Theorem~\ref{thm:dynamic-spanner}) and distance-oracle stretch (Theorem~\ref{inc:oracle}), respectively.

\subsection{Implementation}\label{app:insertion-only-impl}
At the start of each phase, the frozen structures on the spanner $S$ are built directly using Theorems~\ref{thm:mehlhorn} and~\ref{inc:oracle}. Within a phase, processing each edge insertion consists of two steps: updating shortest-path distances in the auxiliary multigraph $L_F$ on the endpoint set $X$, and propagating the resulting candidate-edge weight decreases to the augmented terminal MST $\mathcal M$ on $T_0\cup C_F$.

\paragraph{Maintaining the auxiliary multigraph.}\label{inc:overlay}
Recall that $L_F$ is the auxiliary multigraph on the endpoint set $X$, containing newly inserted edges in $F$ and virtual edges supplied by the distance oracle on $S$. To maintain exact shortest-path distances in $L_F$ while supporting path recovery in time proportional to the number of hops when reporting a Steiner tree, we maintain two matrices on the endpoint set $X$. For each pair $i,j\in X$, the distance matrix stores $D_F(i,j)=d_{L_F}(i,j)$ (abbreviated $D(i,j)$ below). For $i\ne j$, the next-hop matrix $\mathsf{Next}(i,j)$ stores the first edge $ih$ of a shortest path from $i$ to $j$, satisfying
\begin{equation}\label{inc:first-invariant}
D(i,j)=w(ih)+D(h,j).
\end{equation}
For identical endpoints, we set $D(i,i)=0$ and leave $\mathsf{Next}(i,i)$ undefined. Since all edge weights are strictly positive and  $L_F$ is finite, no vertices are repeated and the recovered path must reach its destination.

Suppose an edge $e=uv$ of weight $c$ is inserted. It may introduce zero, one, or two new endpoints into $X$. We ensure that virtual edges connecting $u$ and $v$ to all existing vertices in $X$ are in place before routes crossing $e$ are evaluated by processing each new endpoint first. We compute its distances to existing endpoints in Step~1 and update distances among existing endpoints in Step~2, before inserting $e$ in Step~3. Below, $D_-$ and $D_+$ denote the distance matrices before and after each operation, respectively.

\emph{1) Distances involving a new endpoint.}
Let $z$ be a new endpoint and $X^-$ the endpoint set before adding it. We query and cache the distance estimates $\widehat d_S(z,x)$ for all $x\in X^-$, and add the corresponding virtual edges to $L_F$. For each $i\in X^-$, we set
\begin{equation}\label{inc:activate-row}
D_+(z,i)=D_+(i,z)=\min_{x\in X^-}\bigl\{\widehat d_S(z,x)+D_-(x,i)\bigr\}.
\end{equation}
Indeed, since $z \notin X^-$ and $i \in X^-$, any shortest $z$--$i$ path in $L_F$ must pass through at least one vertex in $X^-$. Let $x \in X^-$ be the first vertex of $X^-$ encountered along the path, splitting it into a $z$--$x$ prefix and an $x$--$i$ suffix. Because every edge inserted earlier has both endpoints in $X^-$, the $z$--$x$ prefix cannot use any newly inserted edge. Thus, with no prior phase insertions affecting $z$--$x$, its $L_F$ distance is given by the virtual edge weight $\widehat{d}_S(z,x)$. Moreover, because $x, i \in X^-$ and the $x$--$i$ suffix does not revisit $z$, its corresponding distance on $L_F$ is $D_-(x,i)$. Since $x$ may be any vertex in $X^-$, minimizing $\widehat{d}_S(z,x) + D_-(x,i)$ over all $x \in X^-$ gives the new distance $D_+(z,i)$ on $L_F$, where $x = i$ covers paths that reach $i$ without visiting any other vertex in $X^-$ since $D_-(i,i) = 0$. Finally, we set $D_+(z,z) = 0$. 

If $x \in X^-$ attains the minimum in \eqref{inc:activate-row}, we set $\mathsf{Next}_+(z,i)$ to the virtual edge $zx$ supplied by the distance oracle on $S$, and in the reverse direction we set $\mathsf{Next}_+(i,z)$ to $\mathsf{Next}_-(i,x)$ if $x\neq i$ and to the virtual edge $iz$ otherwise. Since $|X^-|\le2B$ and each distance-oracle query takes $\Oh(r)$ time, querying the $O(B)$ distance estimates takes $\Oh(rB)$ time. Across all endpoints $i\in X^-$, evaluating the minimum in~\eqref{inc:activate-row} takes $O(B^2)$ time in total. Each corresponding next-hop entry takes constant time to set. Thus this step takes $\Oh(rB+B^2)$ time.

\emph{2) Distances between existing endpoints.}
Adding $z$ may also shorten paths between vertices already in $X^-$. For every $i,j\in X^-$, we set
\begin{equation}\label{inc:activate-old}
D_+(i,j)=\min\bigl\{D_-(i,j),D_+(i,z)+D_+(z,j)\bigr\}.
\end{equation}
A shortest path either avoids $z$, in which case its best possible length is $D_-(i,j)$, or passes through $z$, in which case its length is $D_+(i,z)+D_+(z,j)$. The distances involving $z$ were computed in the previous step. If the route through $z$ is strictly cheaper, we set $\mathsf{Next}_+(i,j)=\mathsf{Next}_+(i,z)$, and otherwise we retain $\mathsf{Next}_-(i,j)$. This step takes $O(B^2)$ time, since updating the distance and next-hop entries for each of the $O(B^2)$ endpoint pairs takes constant time. 

\emph{3) Inserting the edge.}
Once both endpoints are present, we insert $e=uv$ with weight $c$ and set, for every pair $i,j\in X$,
\begin{equation}\label{inc:insert-edge}
\begin{split}
D_+(i,j)=\min\bigl\{D_-(i,j),
D_-(i,u)+c+D_-(v,j),
D_-(i,v)+c+D_-(u,j)\bigr\}.
\end{split}
\end{equation}
A shortest path either avoids $e$ or crosses it in either direction. These are precisely the three expressions above. All entries on the right of \eqref{inc:insert-edge} are read from the matrix obtained after introducing the endpoints but before inserting $e$.

We set $\mathsf{Next}_+(i,j)=\mathsf{Next}_-(i,j)$ if the first expression in~\eqref{inc:insert-edge} achieves the minimum; if the second does, we set $\mathsf{Next}_+(i,j)=\mathsf{Next}_-(i,u)$ when $i\neq u$ and $\mathsf{Next}_+(i,j)=uv$ when $i=u$, and we apply the symmetric rule for the third expression (breaking ties in favor of the earliest expression in the displayed order of~\eqref{inc:insert-edge}). Updating all $O(B^2)$ endpoint pairs takes $O(B^2)$ time, since evaluating the three expressions and setting the corresponding next-hop entry take constant time per pair.

\begin{lemma}\label{inc:overlay-update}
With distance-oracle query time $\Oh(r)$, processing one edge insertion maintains the exact distance and next-hop matrices of $L_F$ in $\Oh(rB+B^2)$ time using $O(B^2)$ space.
\end{lemma}
\begin{proof}
An insertion introduces at most two new endpoints to $X$. Summing the bounds from the three steps above gives total time $\Oh(rB+B^2)$. The distance and next-hop matrices use $O(B^2)$ space; keeping separate matrices for the values before and after each update does not change this bound. Storing $L_F$ also takes $O(B^2)$ space, since it contains at most $B$ inserted edges and one virtual edge supplied by the distance oracle for each pair of endpoints in $X$, where $|X|\le2B$. Thus the total space is $O(B^2)$.
\end{proof}

\subsubsection{Maintaining the MST on $T_0\cup C_F$}
\label{inc:terminal-tree}
Once the updated distances $D_F(x,y)$ in $L_F$ are available, we propagate any improvements to the candidate edges $C_F$ between terminal owners and update the maintained MST $\mathcal M$ of $T_0\cup C_F$. Since $\delta(\cdot)$ is fixed during a phase and insertions only decrease $D_F(x,y)$, candidate weights $c_{xy}=\delta(x)+D_F(x,y)+\delta(y)$ in~\eqref{inc:candidate} are monotonically non-increasing, which allows us to maintain $\mathcal M$ via standard dynamic-tree edge replacements~\cite{ST83}. Specifically, for each unordered pair $x,y\in X$ with distinct owners $s=a(x)$ and $t=a(y)$, we evaluate $c_{xy}=\delta(x)+D_F(x,y)+\delta(y)$, adding the candidate edge to $C_F$ if the pair is new or decreasing its weight otherwise. If a candidate edge with decreased weight already belongs to $\mathcal M$, we update its weight in $\mathcal M$ directly; otherwise, we query the maximum-weight edge on the $s$--$t$ path in $\mathcal M$ and replace it with the candidate edge whenever the candidate edge is strictly lighter, maintaining the edge set of $\mathcal M$ and its total weight $w(\mathcal M)$ explicitly so that cost queries take $O(1)$ worst-case time.

\begin{lemma}\label{lem:inc-terminal-tree}
Maintaining the MST $\mathcal M$ of $T_0\cup C_F$ in a link--cut tree~\cite{ST83} takes $O(|R|\log n)=\Oh(n)$ initialization time at the start of a phase, $O(B^2\log n)=\Oh(B^2)$ amortized time per edge insertion within a phase, and $O(|R|+B^2)$ space.
\end{lemma}
\begin{proof}
A link--cut tree supports finding a maximum-weight edge on a tree path, updating edge weights, and cutting and linking tree edges in $O(\log n)$ amortized time per operation~\cite{ST83}. At the beginning of each phase, we initialize it with $\mathcal M=T_0$ by inserting the $|R|-1$ tree edges using link operations in $O(|R|\log n)=\Oh(n)$ time. During the phase, each candidate-edge insertion or weight decrease requires at most one path-maximum query, one cut, and one link---only a constant number of link--cut tree operations---and therefore takes $O(\log n)$ amortized time. Since there are $O(B^2)$ endpoint pairs, maintaining $\mathcal M$ takes $O(B^2\log n)=\Oh(B^2)$ amortized time per edge insertion. The link--cut tree uses $O(|R|)$ space. Together with $T_0$ and the $O(B^2)$ candidate edges in $C_F$, MST maintenance uses $O(|R|+B^2)$ space.
\end{proof}

\subsection{Approximation Analysis}
\label{inc:approximation}
We now show that $\mathcal M$ provides a $2\sigma A$-approximation throughout each phase. Let $G^\circ$ be the graph at the beginning of the phase and $F$ the set of edges inserted during the phase, so $G=G^\circ\cup F$ is the current underlying graph. Define $H_F=K_S(R)\cup C_F$, where $K_S(R)$ is the terminal metric closure of the frozen spanner snapshot $S$, and $C_F$ contains the candidate edges maintained during the phase. Because any $s$--$t$ shortest path in $G$ that uses newly inserted edges in $F$ must enter and leave the endpoint set $X$ at some first and last vertices $x,y\in X$, its middle subpath between $x$ and $y$ can be approximated in $L_F$ and lifted to terminals $a(x),a(y)\in R$ via $C_F$, while its outer portions in $G^\circ$ are approximated by $K_S(R)$. This gives the following terminal-distance comparison, where $\sigma=(2s-1)(1+\delta)$ is the spanner stretch and $A=2r-1$ is the distance-oracle stretch.

\begin{lemma}[Terminal-distance comparison]\label{inc:sandwich}
For every pair of terminals $s,t\in R$,
\[
d_G(s,t)\le d_{H_F}(s,t)\le\sigma A\,d_G(s,t).
\]
\end{lemma}
\begin{proof}
Every edge of $K_S(R)$ corresponds to a walk in $S \subseteq G$ of equal weight, and each candidate edge in $C_F$ for an endpoint pair $x,y$ represents a walk in $G$ of no greater weight. The predecessor paths from $a(x)$ to $x$ and from $y$ to $a(y)$ have weights $\delta(x)$ and $\delta(y)$, denoting the exact distances from $x$ and $y$ to their respective owners in $S$. Furthermore, any shortest $x$--$y$ path in $L_F$ expands---through direct edges and the path-reporting oracle---into a walk in $G$ of weight at most $D_F(x,y)$. Combining these segments yields a total route weight of at most $c_{xy} = \delta(x) + D_F(x,y) + \delta(y)$. Consequently, every route in $H_F$ admits a walk in $G$ of no greater weight, establishing the lower bound.

For the upper bound, let $P$ be a shortest $s$--$t$ path in $G$. We consider two cases.

\emph{1) The path uses no edge inserted during the phase.}
Then $P$ lies in $G^\circ$, and the metric edge $st$ in $K_S(R)$ has weight
\[
d_S(s,t)\le\sigma d_{G^\circ}(s,t)=\sigma w(P)
\qquad\text{(spanner stretch, Theorem~\ref{thm:dynamic-spanner})}.
\]

\emph{2) The path uses at least one edge inserted during the phase.}
Let $x$ and $y$ be the first and last vertices of the endpoint set $X$ encountered along $P$, respectively, which decomposes $P$ as
\[
P=P_s\circ P_{xy}\circ P_t,
\]
where $P_s$ runs from $s$ to $x$, $P_{xy}$ from $x$ to $y$, and $P_t$ from $y$ to $t$. Both outer portions contain only edges of $G^\circ$: an inserted edge there would have an endpoint in $X$ before $x$ or after $y$.

Let $a=a(x)$ and $b=a(y)$, and consider the walk in $H_F$ given by
\[
s\longrightarrow a\longrightarrow b\longrightarrow t,
\]
where the first and last steps use metric edges of $K_S(R)$, while the middle step uses the candidate edge in $C_F$ associated with $x,y$ (omitting any step between identical terminals). The total weight is at most
\[
d_S(s,a)+\delta(x)+D_F(x,y)+\delta(y)+d_S(b,t).
\]
Eliminating any cycles yields a simple $s$--$t$ path in $H_F$ of no greater weight. We bound the three portions of the displayed expression separately.

For the left portion, since $s$ is a terminal and $a$ is the nearest terminal to $x$ in $S$, we have $\delta(x)=d_S(a,x)\le d_S(s,x)$. By the triangle inequality, $d_S(s,a)\le d_S(s,x)+d_S(x,a)=d_S(s,x)+\delta(x)$. Therefore,
\begin{align*}
d_S(s,a)+\delta(x)
&\le d_S(s,x)+2\delta(x) &&\text{(triangle inequality in $S$)}\\
&\le 3d_S(s,x) &&\text{(nearest-terminal ownership, $\delta(x)\le d_S(s,x)$)}\\
&\le 3\sigma w(P_s) &&\text{(spanner stretch on $P_s\subseteq G^\circ$, Theorem~\ref{thm:dynamic-spanner})}.
\end{align*}
Symmetrically, the right portion satisfies
\[
\delta(y)+d_S(b,t)\le3\sigma w(P_t).
\]

For the middle portion, recall that $D_F(x,y)$ is the exact shortest-path distance in $L_F$, whose edges are either newly inserted edges or virtual edges supplied by the distance oracle on $S$. The path $P_{xy}$ may alternate arbitrarily between old and newly inserted edges. Retaining each newly inserted edge and replacing each consecutive group $Q$ of old edges, with endpoints $u,v$, by the virtual edge $uv\in E(L_F)$ (since $u,v\in X$) gives
\begin{align*}
\widehat d_S(u,v)
&\le A d_S(u,v) &&\text{(oracle stretch, Theorem~\ref{inc:oracle})}\\
&\le A\sigma d_{G^\circ}(u,v)\le A\sigma w(Q) &&\text{(spanner stretch on $Q\subseteq G^\circ$, Theorem~\ref{thm:dynamic-spanner})}.
\end{align*}
Let $\ell_{\rm old}$ and $\ell_{\rm new}$ be the total weights of old and newly inserted edges in $P_{xy}$, respectively. These replacements give an $x$--$y$ route in $L_F$ of weight at most $A\sigma\ell_{\rm old}+\ell_{\rm new}$: replacing old-edge groups incurs a factor of at most $A\sigma$, while newly inserted edges are used directly in $L_F$ with the same weights as in $P_{xy}$. As $D_F(x,y)$ is the shortest distance in $L_F$, it is no greater than the weight of this route. Therefore,
\begin{equation}\label{inc:middle-bound}
D_F(x,y)\le A\sigma w(P_{xy}).
\end{equation}

Adding the bounds gives
\[
\begin{aligned}
d_{H_F}(s,t)
&\le 3\sigma w(P_s)+A\sigma w(P_{xy})+3\sigma w(P_t)\\
&\le \max\{A,3\}\sigma w(P)
=A\sigma d_G(s,t)
\qquad\text{(since $A=2r-1\ge 3$ for $r\ge 2$ and $w(P)=d_G(s,t)$)},
\end{aligned}
\]
which proves the upper bound.
\end{proof}

Lemma~\ref{inc:sandwich} shows that $H_F=K_S(R)\cup C_F$ approximates pairwise terminal distances in $G$ within a factor of $\sigma A$. Combining this distance comparison with the metric-MST bound bounds $\mst(H_F)$ within $2\sigma A\,\OPT_G(R)$, while the MST augmentation property  ensures that $\mathcal M$ achieves the same weight, i.e., $w(\mathcal M)=\mst(H_F)$. Expanding the edges of $\mathcal M$ into their underlying walks in $G$ yields our $2\sigma A$-approximate Steiner tree.

\begin{theorem}\label{inc:phase-approximation}
Expanding the edges of the maintained terminal MST $\mathcal M$ into walks in $G$, taking their union, removing cycles, and pruning nonterminal leaves yields a Steiner tree $T$ for $R$ with
\[
w(T)\le w(\mathcal M)\le2\sigma A\,\OPT_G(R).
\]
Since $\sigma=(2s-1)(1+\delta)$ and $A=2r-1$, we get approximation factor $2(2s-1)(2r-1)(1+\delta)$.
\end{theorem}
\begin{proof}
Let $M_G$ be an MST of the current terminal metric closure $K_G(R)$. For every edge $st\in M_G$, Lemma~\ref{inc:sandwich} gives an $s$--$t$ path $P'_{st}$ in $H_F$ of weight at most $\sigma A d_G(s,t)$. Their union connects all terminals and therefore contains a spanning tree of $H_F$. Consequently,
\begin{align*}
w(\mathcal M)
&=\mst(T_0\cup C_F)=\mst(H_F) &&\text{(MST augmentation, Lemma~\ref{lem:mst-augmentation})}\\
&\le w\left(\bigcup_{st\in E(M_G)}P'_{st}\right) \le\sum_{st\in E(M_G)}w(P'_{st}) &&\text{(the union counts shared edges at most once)}\\
&\le\sigma A\sum_{st\in E(M_G)}d_G(s,t) = \sigma A\,w(M_G) &&\text{(Lemma~\ref{inc:sandwich})}\\
&\le 2\sigma A\,\OPT_G(R) &&\text{(metric-MST bound, Lemma~\ref{lem:metric-mst})}.
\end{align*}
Finally, expanding the edges of $\mathcal M$ into walks in $G$ (and, for the selected candidate edges in $E(\mathcal M)\cap C_F$, replacing the union of their shortest paths in $L_F$ by a spanning forest on $X$ before expanding virtual edges via the path-reporting oracle), taking their union, removing cycles, and pruning nonterminal leaves preserves terminal connectivity without increasing weight, yielding a Steiner tree $T$ with
\[
w(T)\le w(\mathcal M)\le2\sigma A\,\OPT_G(R). \qedhere
\]
\end{proof}

\subsection{Time and Space Analysis}
\label{inc:runtime}
It remains to analyze the running time and space usage for fixed parameters $s,r\ge 2$ and $\delta\in(0,1)$. We first establish the running time of each routine (initialization, phase reconstruction, single-update processing, and tree reporting) as a function of the phase length $B$, and then choose $B$ to balance the update cost. The space usage is bounded in Lemma~\ref{lem:inc-space}.

\begin{lemma}\label{lem:inc-routine-times}
For fixed integers $s,r\ge 2$, constant $\delta\in(0,1)$, and phase length $1\le B\le n$, the procedures of the insertion-only algorithm satisfy the following time bounds:
\begin{enumerate}[nosep,label=(\roman*)]
  \item \emph{Initialization:} $\Oh(m_0+n^{1+1/s+1/r})$ expected time.
  \item \emph{Phase reconstruction:} $\Oh\bigl((n^{1+1/s+1/r})/B + B\bigr)$  expected amortized time per insertion over $B$ insertions.
  \item \emph{Single-update processing:} $\Oh(B^2)$ expected amortized time per insertion.
  \item \emph{Tree reporting:} $\Oh(Bn)$ worst-case time per query.
\end{enumerate}
\end{lemma}
\begin{proof}
We establish the four bounds in order.

\paragraph{Initialization.}
Before the first phase, initializing the dynamic spanner $S_0$ on $m_0$ edges takes $\Oh(n+m_0)$ expected time (Theorem~\ref{thm:dynamic-spanner}). Performing the initial phase reconstruction on the snapshot $S=S_0$ (where $\E[|E(S)|]=\Oh(n^{1+1/s})$ and $L_F,C_F$ are initially empty) takes $\Oh(n^{1+1/s+1/r})$ expected time as shown next, giving $\Oh(m_0+n^{1+1/s+1/r})$ expected initialization time in total.

\paragraph{Phase reconstruction.}
Let $m_S=|E(S)|$ be the number of edges in the frozen spanner snapshot $S$ at the start of a phase, where $m_S\ge n-1$ because $S$ is connected and $\E[m_S]=\Oh(n^{1+1/s})$ by Theorem~\ref{thm:dynamic-spanner}. Copying $S$ and running Mehlhorn's construction to obtain the owner forest and terminal-metric MST $T_0$ takes $\Oh(m_S+n)$ deterministic time (Theorem~\ref{thm:mehlhorn}), building the path-reporting distance oracle on $S$ takes $\Oh(m_S n^{1/r})$ expected time (Theorem~\ref{inc:oracle}), initializing the link--cut tree with $\mathcal M=T_0$ takes $\Oh(n)$ deterministic time (Lemma~\ref{lem:inc-terminal-tree}), and clearing the $O(B^2)$ entries of $L_F$ and $C_F$ takes $O(B^2)$ deterministic time. Taking expectations over $m_S$ and the randomized oracle construction gives expected reconstruction time
\[
\Oh\bigl(\E[m_S]+n+\E[m_S]n^{1/r}+B^2\bigr)=\Oh\bigl(n^{1+1/s+1/r}+B^2\bigr).
\]
Charging each reconstruction after the first to the $B$ insertions of the preceding phase gives
\begin{equation}\label{inc:rebuild-cost}
\Oh\!\left(\frac{n^{1+1/s+1/r}}{B}+B\right)
\end{equation}
expected amortized time per insertion.

\paragraph{Single-update processing.}
On each edge insertion, updating the dynamic spanner $S_t$ takes $\Oh(1)$ expected amortized time (Theorem~\ref{thm:dynamic-spanner}), updating the distance and next-hop matrices of $L_F$ takes $\Oh(rB+B^2)=\Oh(B^2)$ deterministic time (Lemma~\ref{inc:overlay-update}), and maintaining the augmented terminal MST $\mathcal M$ on $T_0\cup C_F$ takes $\Oh(B^2)$ deterministic amortized time (Lemma~\ref{lem:inc-terminal-tree}). Summing these costs gives expected amortized single-update time $\Oh(B^2)$.

\paragraph{Tree reporting.}
To materialize an explicit Steiner tree on query, we first partition the $|R|-1$ edges of $\mathcal M$ into $E(\mathcal M)\cap E(T_0)$ and $E(\mathcal M)\cap C_F$ in $O(|R|)=O(n)$ time. Because every candidate edge in $C_F$ connects two terminals in the active owner set $A_F=\{a(x):x\in X\}\subseteq R$ of size $|A_F|\le |X|\le 2B$ and $\mathcal M$ is acyclic, the subgraph $E(\mathcal M)\cap C_F$ is a forest on $A_F$ and therefore contains at most $|A_F|-1<2B$ candidate edges. Next, we include the stored cross-region edges in $S$ for $E(\mathcal M)\cap E(T_0)$ and coherently expand all required owner-forest routes---both from the cross-region endpoints of $E(\mathcal M)\cap E(T_0)$ and from the endpoints $x,y\in X$ associated with $E(\mathcal M)\cap C_F$---by walking up the rooted owner forest in $S$, marking visited vertices and stopping each walk upon reaching a terminal root or an already marked vertex (as in Lemma~\ref{lem:coherent-expansion}). This traverses each of the at most $n-1$ owner-forest edges at most once, taking $O(n)$ time and contributing $O(n)$ edges. For each of the fewer than $2B$ candidate edges in $E(\mathcal M)\cap C_F$ with endpoint pair $(x,y)\in X^2$, we recover a shortest $x$--$y$ path $P_{xy}$ in $L_F$ of at most $|X|-1<2B$ hops via $\mathsf{Next}$ in $O(B^2)$ total time, and extract a spanning forest $F_L\subseteq\bigcup P_{xy}$ on $X$ in $O(B^2)$ time. Since $F_L$ is a forest on $X$, it connects every such pair $(x,y)$ with weight $w_{L_F}(F_L)\le\sum D_F(x,y)$ and contains at most $|X|-1<2B$ edges of $L_F$. Each edge of $F_L$ is either a newly inserted edge in $F$ or a virtual edge expanded into a simple path $P$ in $S$ of at most $n-1$ edges by the path-reporting oracle in $\Oh(r+|E(P)|)=\Oh(n)$ time (Theorem~\ref{inc:oracle}), so expanding $F_L$ requires fewer than $2B$ oracle queries and produces $O(Bn)$ edge occurrences in $\Oh(Bn)$ time. Finally, removing cycles and pruning non-terminal leaves across the $O(Bn)$ collected edges takes $O(Bn)$ time, giving $\Oh(Bn)$ worst-case reporting time.
\end{proof}
Combining the single-update cost with the amortized phase-reconstruction cost shows that any sequence of $U\ge 1$ insertions takes $\Oh(UB^2+\lfloor U/B\rfloor n^{1+1/s+1/r})$ expected time beyond initialization, corresponding to an expected amortized update time of
\begin{equation}\label{inc:general-runtime}
T_{\mathrm{upd}}(B)
=\Oh\!\left(\frac{n^{1+1/s+1/r}}{B}+B\right)+\Oh(B^2)=\Oh\!\left(\frac{n^{1+1/s+1/r}}{B}+B^2\right).
\end{equation}
Choosing
$B=\left\lceil n^{(1+1/s+1/r)/3}\right\rceil$
yields expected amortized update time
$T_{\mathrm{upd}}=\Oh\!\left(n^{\frac23(1+1/s+1/r)}\right)$,
and tree-reporting time $\Oh(Bn)=\Oh(n^{4/3+(1/s+1/r)/3})$.

\begin{lemma}\label{lem:inc-space}
Let $m_t=|E(G_t)|$ be the current edge count. For fixed integers $s,r\ge 2$ with $1/s,1/r\le 1/2$ and our choice of $B$, the total expected space used is
\[
\Oh\!\left(m_t+n^{1+1/\min\{s,r\}}\right).
\]
\end{lemma}
\begin{proof}
Storing the current graph takes $O(m_t+n)$ space; maintaining the dynamic spanner $S_t$ and its frozen snapshot $S$ takes $\Oh(n^{1+1/s})$ expected space (Theorem~\ref{thm:dynamic-spanner}); storing the path-reporting distance oracle on $S$ takes $\Oh(n^{1+1/r})$ expected space (Theorem~\ref{inc:oracle}); and storing the owner forest and $T_0$ takes $O(n)$ space (Theorem~\ref{thm:mehlhorn}). Within a phase, maintaining $L_F$ takes $O(B^2)$ space (Lemma~\ref{inc:overlay-update}) and maintaining $\mathcal M$ on $T_0\cup C_F$ takes $O(|R|+B^2)$ space (Lemma~\ref{lem:inc-terminal-tree}). Summing these bounds and using $|R|\le n$ gives total expected space
\begin{equation*}
\Oh\!\left(m_t+n^{1+1/s}+n^{1+1/r}+B^2\right).
\end{equation*}
For $B=\lceil n^{(1+1/s+1/r)/3}\rceil$ and $a=\max\{1/s,1/r\}=1/\min\{s,r\}\le 1/2$, we have $\frac23(1+1/s+1/r)\le\frac23(1+2a)\le 1+a$, so $B^2=O(n^{1+a})$ is absorbed by $\Oh(n^{1+1/s}+n^{1+1/r})=\Oh(n^{1+1/\min\{s,r\}})$.\qedhere
\end{proof}

Combining Theorem~\ref{inc:phase-approximation} with Lemmas~\ref{lem:inc-routine-times} and~\ref{lem:inc-space} yields the full parameterized guarantee for our insertion-only data structure.

\begin{theorem}[Parameterized version of Theorem~\ref{thm:main-incremental}]\label{inc:main}
Let $G$ be a connected $n$-vertex incremental undirected graph with $m_0$ edges initially, a terminal set $R$, and polynomially bounded positive edge weights over all updates. For any fixed integers $s,r\ge 2$ and constant $0<\delta<1$, setting phase length $B=\bigl\lceil n^{(1+1/s+1/r)/3}\bigr\rceil$ gives a randomized insertion-only data structure against an oblivious adversary that maintains a $2(2s-1)(2r-1)(1+\delta)$-approximate Steiner tree for $R$ in $G$ with:
\begin{itemize}[nosep]
  \item $\Oh\!\left(n^{\frac23(1+1/s+1/r)}\right)$ expected amortized update time,
  \item $\Oh\!\left(m_0+n^{1+1/s+1/r}\right)$ expected initialization time, and
  \item $\Oh\!\left(m_t+n^{1+1/\min\{s,r\}}\right)$ expected space, where $m_t$ is the current number of edges.
\end{itemize}
After every insertion, the approximation guarantee holds deterministically, cost queries take $O(1)$ worst-case time, and reporting an explicit Steiner tree takes $\Oh\!\left(n^{4/3+(1/s+1/r)/3}\right)$ worst-case time.
\end{theorem}
\begin{proof}
Approximation follows from Theorem~\ref{inc:phase-approximation}, returning the maintained scalar weight $w(\mathcal M)$ on cost queries in $O(1)$ worst-case time. Initialization, amortized update, and worst-case reporting times follow from Lemma~\ref{lem:inc-routine-times} and~\eqref{inc:general-runtime} (since $1\le B=\lceil n^{(1+1/s+1/r)/3}\rceil\le\lceil n^{2/3}\rceil\le n$ for all $s,r\ge 2$ and $n\ge 1$), and expected space follows from Lemma~\ref{lem:inc-space}.

In particular, for any fixed $0<\zeta<1/3$, choosing $s=r=\lceil 2/\zeta\rceil$ gives $1/s=1/r\le\zeta/2$ and $(2s-1)(2r-1)=O(1/\zeta^2)$, establishing Theorem~\ref{thm:main-incremental} (with update time $\Oh(n^{2/3+2\zeta/3})$, space $\Oh(m_t+n^{1+\zeta/2})$, and reporting time $\Oh(n^{4/3+\zeta/3})$). Choosing $s=r=5$, $\delta=\varepsilon/162$, and $B=\lceil n^{7/15}\rceil$ yields a $(162+\varepsilon)$-approximation in $\Oh(n^{14/15})$ expected amortized update time.
\end{proof}

\section*{AI Use Statement.}
We used GPT-5.6-Sol to learn about the problem and explore candidate constructions. While the main ideas behind the constructions and abstractions are our own, GPT-5.6-Sol was invaluable in exploring proof approaches and in formalizing proof sketches. The authors verified the correctness and originality of all content, including references.
\bibliographystyle{alpha}
\bibliography{references}
\end{document}